\documentclass[11pt]{article}
\usepackage{epsfig}
\usepackage{lscape}
\usepackage{longtable}

\usepackage{subfigure}  
\usepackage{subfig}  
\usepackage{color}
\usepackage{multirow}
\usepackage{epsfig}
\usepackage{wrapfig}
\usepackage{rotating}
\usepackage{amsmath}
\usepackage[greek,english]{babel} 
\usepackage{setspace} 
\usepackage{eurosym}
\usepackage{graphics}
\usepackage{wasysym}

\usepackage[T1]{fontenc}
\usepackage[english]{babel}
\usepackage{amsmath}
\usepackage{amssymb,amsfonts,textcomp}
\usepackage{array}
\usepackage{supertabular}
\usepackage{hhline}

\newcommand{\be}{\begin{eqnarray}}
\newcommand{\ee}{\end{eqnarray}}
\def\lsim{\mathrel{\rlap{\lower4pt\hbox{\hskip 0.5 pt$\sim$}}
\raise1pt\hbox{$<$}}}                
\def\gsim{\mathrel{\rlap{\lower4pt\hbox{\hskip1pt$\sim$}}
\raise1pt\hbox{$>$}}} 
    
\newcommand{\s}{{\rm s}}
\newcommand{\m}{{\rm m}}
\newcommand{\apr}{{A^\prime}}
\newcommand{\MeV}{{\rm MeV}}
\newcommand{\GeV}{{\rm GeV}}

\makeatletter
\newcommand\arraybslash{\let\\\@arraycr}
\makeatother
\newcommand{\thedate}{\today}

\input epsf
\def\beqn{\begin{eqnarray}}
\def\eeqn{\end{eqnarray}}
\def\beq{\begin{equation}}
\def\eeq{\end{equation}}

\def\gevc2{(GeV/c)$^2$}
\newcommand*{\jlab}{Jefferson Lab, Newport News, VA 23606, USA}
\newcommand*{\nhs}{University of New Hampshire, Durham NH 03824, USA}
\newcommand*{\perimeter}{Perimeter Institute for Theoretical Physics, Waterloo, Ontario, Canada, N2L 2Y5}

\newcommand*{\frascati}{Istituto Nazionale di Fisica Nucleare, Laboratori Nazionali di Frascati, P.O. 13, 00044 Frascati, Italy}
\newcommand*{\genova}{Istituto Nazionale di Fisica Nucleare, Sezione di Genova\\ e Dipartimento di Fisica dell'Universit\`a, 16146 Genova, Italy}
\newcommand*{\sanita}{Istituto Nazionale di Fisica Nucleare, Sezione di Roma e Gruppo Collegato Sanit\`a,
 e  Universit\`a La Sapienza, Italy}

\newcommand*{\infnba}{Istituto Nazionale di Fisica Nucleare, Sezione di Bari e Dipartimento di Fisica dell'Universit\`a, Bari, Italy}
\newcommand*{\infnfe}{Istituto Nazionale di Fisica Nucleare, Sezione di Ferrara e Dipartimento di Fisica dell'Universit\`a, Ferrara, Italy}

\newcommand*{\ohio}{Ohio University, Department of Physics, Athens, OH 45701, USA}
\newcommand*{\gwu} {The George Washington University, Washington, D.C., 20052}

\newcommand*{\odu}{Old Dominion University, Department of Physics,
Norfolk VA 23529, USA}

\newcommand*{\edinb}{Edinburgh University, Edinburgh EH9 3JZ, United Kingdom}
\newcommand*{\glasgow}{University of Glasgow, Glasgow G12 8QQ, United Kingdom}

\newcommand{\sassari}{Universit\`a di Sassari e Istituto Nazionale di Fisica Nucleare, 07100 Sassari, Italy}
\newcommand{\torvergata} {Istituto Nazionale di Fisica Nucleare, Sezione di Roma-TorVergata e Dipartimento di Fisica dell'Universit\`a, Roma, Italy}
\newcommand{\catania} {Istituto Nazionale di Fisica Nucleare, Sezione di Catania,  Catania, Italy}
\newcommand{\torino} {Istituto Nazionale di Fisica Nucleare, Sezione di Torino,  Torino, Italy}
\newcommand{\padova} {Istituto Nazionale di Fisica Nucleare, Sezione di Padova,  Padova, Italy}
\newcommand*{\hu}{Department of Physics, Hampton University, Hampton VA 23668, USA }

\begin{document}

\begin{center}
{\tiny \leftline{V1.8}}
{\tiny\leftline{\thedate}}
\date{\today}
\rightline{Letter of Intent to PAC 42}
\vskip 1.0cm
{\bf\huge Dark matter search in a Beam-Dump eXperiment (BDX) at Jefferson Lab}
\vskip 1.cm
{  \large \it The BDX Collaboration }
\vskip 0.5cm

{M.~Battaglieri\footnote{Contact Person, email: Marco.Battaglieri@ge.infn.it}\footnote{Spokesperson}, A.~Bersani, A.~Celentano$^\dag$, R.~De~Vita$^\dag$,  E.~Fanchini, S.~Fegan, P.~Musico, M.~Osipenko, M.~Ripani, E.~Santopinto, M.~Taiuti\\}
{\small\it\genova}
\bigskip

{E. Izaguirre$^\dag$, G. Krnjaic$^\dag$, P. Schuster, N. Toro\\}
{\small\it\perimeter}
\bigskip

{M. Dalton, A. Freyberger, F.-X.~ Girod, V. Kubarovsky,
 E. Smith$^\dag$, S. Stepanyan$^\dag$, M. Ungaro\\}
{\it\small\jlab}
\bigskip

{G.~De~Cataldo, R.~De~Leo, D.~Di~Bari, L.~Lagamba, E.~Nappi, R.~Perrino\\}
{\small\it \infnba}
\bigskip

{M.~Carpinelli, V.~Sipala\\}
{\small\it\sassari}
\bigskip

{S.~Aiello, V.~Bellini, M.~De Napoli, A.~Giusa, F.~Mammoliti, E.~Leonora, F.~Noto, N.~Randazzo, G.~Russo, M.~Sperduto, C.~Sutera, C.~Ventura\\}
{\small\it\catania\\}
\bigskip

{L.~Barion, G.~Ciullo, M.~Contalbrigo, P.~Lenisa, A.~Movsisyan, F.~Spizzo, M.~Turisini\\}
{\small\it \infnfe}
\bigskip

{F.~De~Persio, E.~Cisbani, C.~Fanelli, F.~Garibaldi, F.~Meddi, G.~M.~Urciuoli\\}
{\small\it \sanita}
\bigskip

{S.~Anefalos Pereira, E.~De~Sanctis, D.~Hasch, V.~ Lucherini, M.~Mirazita, R.~Montgomery, S.~Pisano\\}
{\small\it \frascati}
\bigskip

{G.~Simi\\}
{\small\it\padova\\}
\bigskip

{ A.~D'Angelo, L.~Colaneri L.~Lanza, A.~Rizzo, C.~Schaerf, I.~Zonta \\}
{\small\it \torvergata}
\bigskip

{D.~Calvo, A.~Filippi\\}
{\small\it\torino}
\bigskip

{M.Holtrop, R.~Peremuzyan\\}
{\it\small\nhs}
\bigskip

{D. Glazier, D.~Ireland, B.~McKinnon, D. Sokhan\\}
{\it\small\glasgow}
\bigskip

{A.~Afanasev, B.~Briscoe\\}
{\it\small\gwu}
\bigskip

{N.~Kalantarians\\}
{\it\small\hu}
\bigskip

{L.~El~Fassi, L.~Weinstein\\}
{\it\small\odu}
\bigskip

{P.~Beltrame, A.~Murphy, D. Watts, L.~Zana\\}
{\it\small\edinb}
\bigskip

{K.~Hicks\\}
{\it\small\ohio}
\bigskip

\begin{abstract}
MeV-GeV dark matter (DM) is theoretically well motivated but remarkably unexplored.  
This Letter of Intent presents the MeV-GeV DM discovery potential for a 1 m$^3$ segmented plastic scintillator detector placed downstream of the beam-dump at one of the high intensity JLab experimental Halls, receiving up to 10$^{22}$ electrons-on-target (EOT) in a one-year period. This experiment (Beam-Dump eXperiment or BDX)  is sensitive to DM-nucleon elastic scattering at the level of a thousand counts per year, with very low threshold recoil energies ($\sim$1 MeV), and limited only by reducible cosmogenic backgrounds. Sensitivity to DM-electron elastic scattering and/or inelastic DM would be below 10 counts per year after requiring all electromagnetic showers in the detector to exceed a few-hundred MeV, which dramatically reduces or altogether eliminates all backgrounds. Detailed Monte Carlo simulations are in progress to finalize the detector design and  experimental set up. An existing 0.036 m$^3$ prototype based on the same technology will be used to validate simulations with background rate estimates, driving the necessary R$\&$D towards an optimized detector. The final detector design and experimental set up will be presented in a full proposal to be submitted to the next JLab PAC. A fully realized experiment would be sensitive to large regions of DM parameter space, exceeding the discovery potential of existing and planned experiments by two orders of magnitude in the MeV-GeV DM mass range.

\end{abstract}

\vskip 1.0cm

\end{center} 

\newpage
\tableofcontents\newpage

\section{Overview}  
In this letter 
we present the potential of a beam-dump experiment to search for light (MeV-GeV) Dark Matter (DM). DM in this mass range is motivated by both experimental and theoretical considerations. Simple extensions to the Standard Model (SM) can accommodate DM-SM interactions that yield the observed DM cosmological abundance. Such models also  generically feature particles
 that explain the currently discrepant value of the muon's anomalous magnetic moment and resolve anomalies in astrophysical observations, while simultaneously evading
 cosmological and direct-production constraints. The physics motivations for light DM are presented in detail in Sec.~\ref{sec:theory}.
  
This experiment could be performed by placing a detector downstream of one of the JLab experimental Halls to measure DM particles that could be
produced by the electron beam in the dump, pass through surrounding shielding material, and deposit visible energy inside the detector by 
scattering off various target particles or (if unstable) by decaying inside the detector volume. 
The run would be completely parasitic without affecting the normal operations and the physics program of the Hall. 
To detect the small signal produced by a light DM particle (mass of tens of MeV)  scattering off a nucleon, the detection 
thresholds need to be fixed at values as low as possible ($\sim$ MeV). 
On the other hand, low thresholds can introduce spurious signals from beam-related (neutrinos) and cosmogenic background (muons, 
neutrons and neutrinos), limiting the measurement  sensitivity. 
Other processes such as elastic scattering off atomic electrons and inelastic
 DM interaction lead to scattered  electrons with energy 
of few GeV that could be easily detected over a null background.
Due to the sensitivity to these different DM signals,
BDX exceeds the sensitivity of other experiments 
 proposed at FNAL, CERN, and LNF. 

The goal of the BDX Collaboration is to propose a search for light DM particles produced with electron beams. This experimental approach is 
capable of covering new theoretically motivated ground in the parameter space of light DM.
The design and simulation of such a facility at Jefferson Lab is not yet complete and this
LOI presents the studies to date that will serve as a basis for a complete proposal. To achieve the desired sensitivity, the ultimate 
experiment must be mounted behind a high-intensity hall, likely Hall A. However, the sensitivity of the  experiment to 
signal and capability to reject backgrounds in this LOI is obtained by scaling up the results of a detailed case study based 
on an existing detector, CORMORINO. 

The  CORMORINO case study was performed under the assumption that the detector would be 
located downstream of the Hall-D beam-dump. The initial intent was to use that configuration as a practical intermediate stepping stone 
by exploiting the inexpensive opportunity offered to us by the above-ground Hall D beam-dump. This 
case study is described in detail here, as it represents a fairly complete simulation of backgrounds and challenges faced by the final experiment.
Results of GEANT4 Monte Carlo simulations to describe  the effect of the beam-dump and the detector response are reported 
in Sec.~\ref{sec:expsetup}. Based on the same simulations, results of  a detailed study 
of processes contributing to the backgrounds with threshold down to 1 MeV are reported in the succeeding subsections.  
Projections, counting rates for signal and background scaled to 
the full experiment, and the expected reach of BDX 
compared to similar experiments  are reported in the last Sec.~\ref{sec:fullexp}. 

The outcome of the case  study is that the dominant backgrounds to a beam dump experiment at a high-duty-cycle machine are not
beam related. Therefore, tests to determine the capabilities of a detector to reject backgrounds need not be performed in
the vicinity of an electron beam dump, and can be optimized at a more convenient location.  
We plan to fully  characterize and measure  the cosmogenic background with CORMORINO  to validate simulations and 
optimize the location and the detector design  for the full-scale experiment. The planned R$\&$D, as well as a list of possible topics 
we want to investigate,  is reviewed in the last Section.

The goal of this LOI is to request guidance from the PAC on our continued effort to develop a full 
proposal at a later date. First, we seek endorsement of the physics goals of this research project. 
Second, we welcome input on interfaces with facility infrastructure that are required near 
existing experimental areas and comments on our program to determine the best detector for the DM search.

\section{Direct dark matter search in  beam-dump experiments}

\subsection{Physics motivations}\label{sec:theory}

Although overwhelming astrophysical and cosmological evidence supports the existence of DM, its elementary properties remain largely elusive.  
The expectation that dark matter has some kind of interaction with Standard Model matter is strongly motivated by two possible mechanisms to explain its origin -- either as an annihilation product of thermal Standard Model matter in the early Universe, or as a product of the same unknown processes that generate the baryon asymmetry. 
There is currently an active program to probe particle DM scattering with direct detection experiments, annihilation with indirect detection telescopes, and
production with particle accelerators.  However, most of these efforts are designed to find heavy (10$-$1000 GeV) DM candidates and sharply lose sensitivity to
lighter (sub-GeV) states whose signals are either too feeble to detect or lie in high-background regions.

In fact, dark matter candidates are readily motivated in the entire MeV-to-TeV range.
Much heavier dark matter is disfavoured because its naive thermal abundance exceeds the observed cosmological matter density.
Much beneath an MeV, astrophysical and cosmological constraints allow only dark matter with ultra-weak couplings to quarks and leptons \cite{Essig:2013lka}.
 Between these boundaries (MeV - TeV), simple models of dark matter can account for its observed abundance through either thermal freeze-out or non-thermal mechanisms.  
The importance of broadening the experimental search program to include the MeV-GeV mass range is underscored by the lack of evidence for 
weak-scale ($\gtrsim 100$ GeV) dark matter scattering through $Z$ bosons, ever stronger constraints on Higgs mediated scattering, and 
by the absence to date of evidence for new SM-charged matter at the LHC. 
 
The status of experimental searches for MeV--TeV dark matter can be summarized as follows.  
The best constraints on multi-GeV dark matter interactions are from underground searches for nuclei recoiling off non-relativistic dark matter particles in the Galactic halo. However these searches are insensitive to few-GeV or lighter dark matter, whose nuclear scattering transfers  invisibly small kinetic energy to a recoiling nucleus.  Direct detection using electron-scattering offers an alternative strategy to search for sub-GeV dark matter, but with dramatically higher backgrounds \cite{Essig:2012yx,Essig:2011nj,Graham:2012su}. Among the best-motivated models of MeV-to-GeV-mass dark matter are those whose interactions with ordinary matter are mediated by new GeV-scale  {\it`dark'} force carriers (for example, a gauge boson that kinetically mixes with the photon) \cite{Boehm:2002yz,Boehm:2003hm,Pospelov:2007mp,ArkaniHamed:2008qn}.  Such models readily account for the stability of dark matter and its observed relic density are compatible with all observations, and have important implications beyond the dark matter itself.  
In these scenarios, dark matter production at high-energy accelerators is generically buried under QCD background, making collider searches for light dark matter insensitive. 

\begin{figure}[t!]
\center
\includegraphics[width=8.3cm]{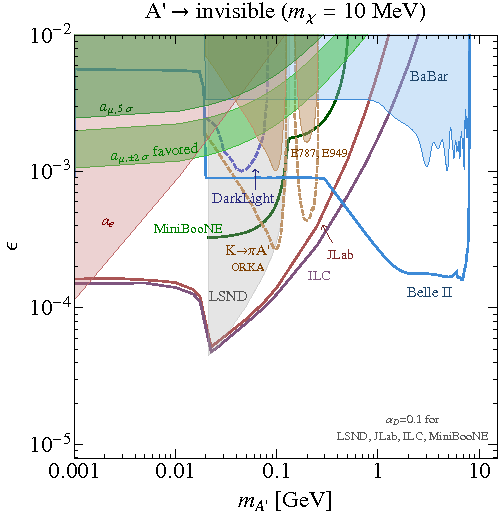}
\caption{
The international effort to search for GeV-scale dark matter and invisibly decaying vector mediators is partially summarized in the Report of the Community Summer Study 2013 (Snowmass) "New, Light, Weakly-Coupled Particles" subgroup \cite{Essig:2013lka}, with additional references therein. In the above plot, representative sensitivity of several approaches is illustrated in the parameter space of mediator coupling $\epsilon$ vs. mediator mass $m_{A'}$ for invisible decays of the mediator into dark matter candidates. The approaches include rare meson decay studies, proton and electron beam fixed target and B-factory collider searches. 
Shaded regions denote constraints from past experiments (LSND, muon and electron $g-2$, and BaBar) and the region favoured by the muon $g-2$ anomaly (lower green band); dashed lines correspond to potential sensitivities for proposed/future experiments.   
Importantly, the approaches shown {\it cannot} be compared model-independently, and this plot is only representative of a particular parameter point in the kinetic mixing model of \cite{Essig:2013lka}.}
\label{fig:Snowmass}
\end{figure}

Dark matter in the $\MeV-\GeV$ range is therefore still relatively unexplored, and has been a topic of considerable recent interest (see e.g. \cite{Essig:2013lka} and references therein, and \cite{Dharmapalan:2012xp,Diamond:2013oda,Izaguirre:2014dua,Morrissey:2014yma}). The sensitivity of past, ongoing, and proposed experiments to dark sector scenarios where dark matter interacts with the Standard Model through a kinetically mixed vector boson, $A'$, is illustrated in Figure \ref{fig:Snowmass}, originally from \cite{Essig:2013lka} (we note that the sensitivity for a ``generic'' JLAB experiment anticipated in \cite{Essig:2013lka} is somewhat weaker, depending on $A'$ mass, than the sensitivity of the experiment proposed in this Letter of Intent and presented in the last section).  While this figure illustrates the power and complementarity of the various experiments, it is of limited use for comparing experiments that use different search techniques, whose sensitivities change considerably for different parameter choices or model assumptions.

The considerable sensitivity of beam-dump experiments to light dark matter is underscored by the reach of existing neutrino experiments \cite{Batell:2009di,deNiverville:2011it, deNiverville:2012ij,Dharmapalan:2012xp,Essig:2013lka}. 
For example, the LSND measurement of electron-neutrino scattering \cite{Auerbach:2001wg} can be used to derive the most stringent constraints to date on the parameter space for invisibly-decaying dark mediators that couple to both baryons and leptons \cite{deNiverville:2011it}. That experiment delivered $\sim 10^{23}$ 800 MeV protons to the LANSCE beam-dump. 
For very low mass $A'$s and dark matter sufficiently light ($100 \MeV \lesssim m_{A'}\lesssim 2 m_{\chi}$), the produced neutral pions have a small exotic decay into $A^\prime$s which 
then decay to $\chi$. The $\chi$ can then scatter off electrons in the LSND detector via $A'$-exchange.
However, the sensitivity of LSND vanishes if the mediator couples only to leptons or baryons and is weakened if its coupling to either is suppressed. 
Relatedly, MiniBooNE is currently being used to search for light dark matter in a dedicated run of the BooNE proton beam at FNAL \cite{MiniBoone2013}. 

Recently it was shown that electron-beam fixed target experiments offer powerful sensitivity to a broad class of dark sector scenarios with particle dark matter in the $\MeV-\GeV$ mass range \cite{Izaguirre:2013uxa,Diamond:2013oda,Izaguirre:2014dua}. Electron beam-dump experiments are complementary to dedicated efforts at proton beam facilities, have comparable DM scattering yield, 
can run parasitically and on a smaller scale than proton-beam counterparts,
and benefit from negligible beam-related backgrounds. 
Such searches can dramatically improve sensitivity to MeV-to-GeV mass dark matter and other long-lived weakly coupled particles, extending well beyond the reach of proposed neutrino-factory experiments and Belle-II projections.
The power of electron beam dump experiments in this context is illustrated
by the existing sensitivity of the SLAC E137 experiment \cite{Bjorken:1988as}. That experiment was sensitive to invisibly decaying dark mediators produced in fixed target collisions involving 20 GeV electrons and the E137 beam-dump \cite{EssigSurujon}.
Despite the rather high energy threshold ($\sim$ 3 GeV) required to see secondary scattering of dark matter particles off electrons, and the small geometric acceptance,
E137  has already probed mediator mixings beyond that probed by proton beam-dumps at intermediate masses. 
In a year of parasitic running, BDX will receive roughly 100 times the charge deposited on E137, with a comparable solid angle, higher-density detector, and lower energy threshold.

\begin{figure}[t!]
\center
\includegraphics[width=8.3cm]{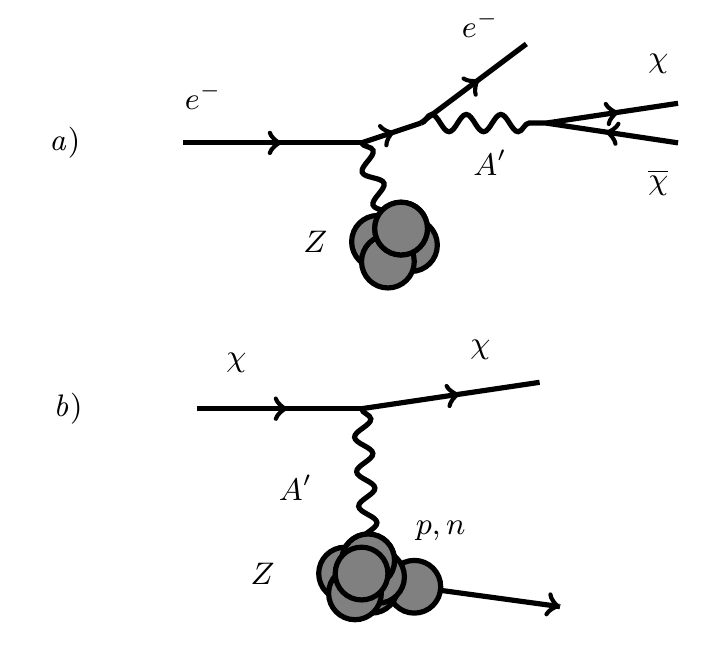}
\caption{  a) $\chi \bar \chi$ pair production in electron-nucleus collisions via the Cabibbo-Parisi radiative process (with $A'$ on- or off-shell) and b) 
$\chi$  scattering off a detector nucleus and liberating a constituent nucleon. 
For the momentum transfers of interest, the incoming $\chi$ resolves the nuclear substructure, so the typical reaction is quasi-elastic and 
nucleons will be ejected.
 }\label{fig:prod}
\end{figure}

\subsection{Light dark datter models and theory of production and detection}\label{sec:elastic}
Whether the dark sector is quite simple or has a rich sector of light particles, the  fixed-target phenomenology of stable $\chi$s (or unstable $\chi$s with lab-frame lifetimes $\gtrsim \mu\s$) is usually well-described by the simplest case --- a renormalizable $U(1)_D$ dark sector with a single stable matter particle $\chi$. For fermionic $\chi$  and $U(1)_D$ coupling to Standard Model matter via kinetic mixing \footnote{We focus in this letter on fermionic $\chi$, but the same approach is sensitive to scalar $\chi$ as well.  Generalizations to $U(1)_D$ coupling to baryonic or leptonic currents have more dramatic effects on the phenomenology.},
 \be
 \label{eq:lagrangian}
{\cal L}_{dark} &=& 
-\frac{1}{4}F^\prime_{\mu\nu} F^{\prime\,\mu\nu} + \frac{\epsilon_Y}{2} F^\prime_{\mu\nu} B_{\mu \nu} + \frac{m^2_{A^\prime}}{2} A^{\prime}_\mu A^{\prime\, \mu} +  \bar \chi ( i \displaystyle{\not}{D}- m_\chi) \chi,
\ee
where $B_{\mu \nu} = B_{[\mu,\nu]}$ and  $F^\prime_{\mu \nu} =
 A^\prime_{[\mu,\nu]}$ are respectively the  hypercharge and dark-photon field strengths and $D_\mu  = \partial_\mu + i g_D A^\prime_\mu$ (and similarly for scalar $\chi$).

 The kinetic mixing parameter $\epsilon_Y$ can arise generically from loops of heavy particles charged under both hypercharge and $U(1)_D$
 and is naturally small, on the scale of $\frac{e g_D}{16\pi^2} \log(M/\Lambda) \sim 10^{-5}-10^{-2}$, where $M$ is the mass of the particle in the loop and $\Lambda$ the theory cutoff scale.
It will be convenient to take as a free parameter $\epsilon\equiv \epsilon_Y cos \theta_W$ (where $\theta_W$ is the weak mixing angle), rather than $\epsilon_Y$ itself.

Upon diagonalizing the kinetic mixing terms in \eqref{eq:lagrangian}, ordinary electrically charged matter acquires a ``dark millicharge'' coupling to the $A^\prime$ of strength $\epsilon e$, while the $\chi$ remains electrically neutral. As a consequence, long-lived dark sector particles $\chi$ couple to ordinary matter primarily through $A'$ exchange.

In this theory, $\chi$s can therefore be pair-produced radiatively in electron-nucleus collisions in the dump (see Fig.~\ref{fig:prod}a).  A fraction of these relativistic particles then scatter off nucleons, nuclei, or electrons in the detector volume (see Fig.~\ref{fig:prod}b).

If $m_{A^\prime} < 2 m_\chi$,  the dominant $\chi$ production mechanism in
an electron fixed-target experiment is the radiative process illustrated in 
Fig. \ref{fig:prod}a) with off-shell $A'$. In this regime, the $\chi$ production yield scales as $\sim \alpha_D \epsilon^2/m^2_\chi$ ($\alpha_D\equiv{g_D}^2/4\pi$),
while $\chi$-nucleon scattering in the detector via $A^{\prime}$ exchange (see Fig. \ref{fig:prod}b))
 occurs with a rate proportional to $ \alpha_D  \epsilon^2/m_{A'}^2$ over most of the mass range. 
 Thus, the total signal yield scales as
\be
N_{\chi} \sim   \frac{  \alpha_D^2 \epsilon^4 }{m_{\chi}^2m_{A'}^2}.  ~~
\ee

If $m_{A^\prime} > 2 m_\chi$, the secondary $\chi$-beam arises from
 radiative $A^{\prime}$ production followed by $A^\prime \to \bar \chi \chi$ decay.  In this regime, the $\chi$ production and
  the detector scattering rates are respectively proportional to $\epsilon^2/m_{A'}^2$  and $  \alpha_D  \epsilon^2/m_{A'}^2$ and the signal yield scales as  
\be
N_{\chi} \sim  \frac{ \alpha_D \epsilon^4}{m_{A'}^4} ~~.
\ee
Thus, for each $\alpha_D$ and $m_{A'}$, we can extract an $\epsilon$-sensitivity corresponding to a given scattering yield.
\begin{figure}[t!] 
\center
\includegraphics[width=0.4785\textwidth]{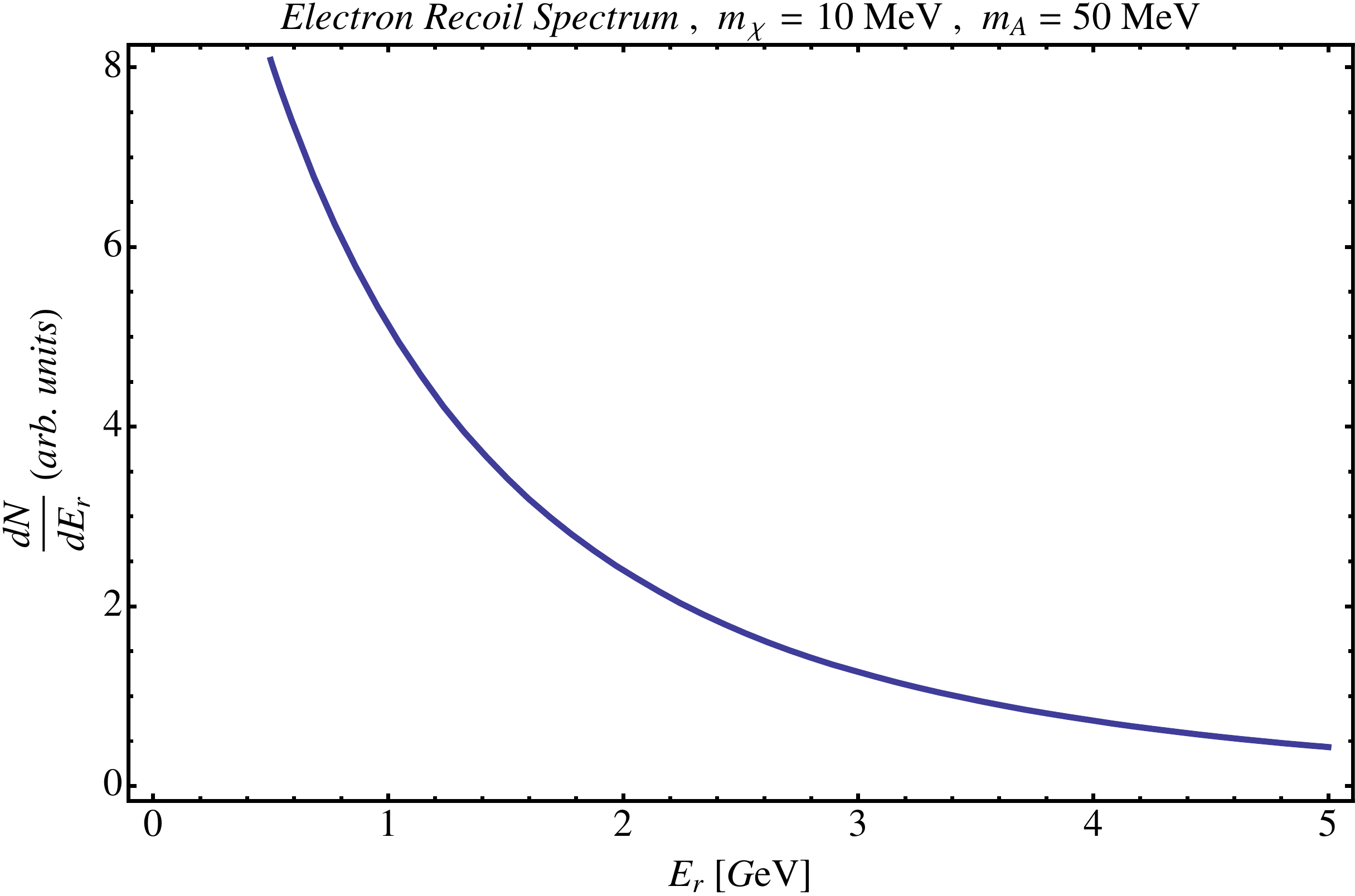} 
\includegraphics[width=0.485\textwidth]{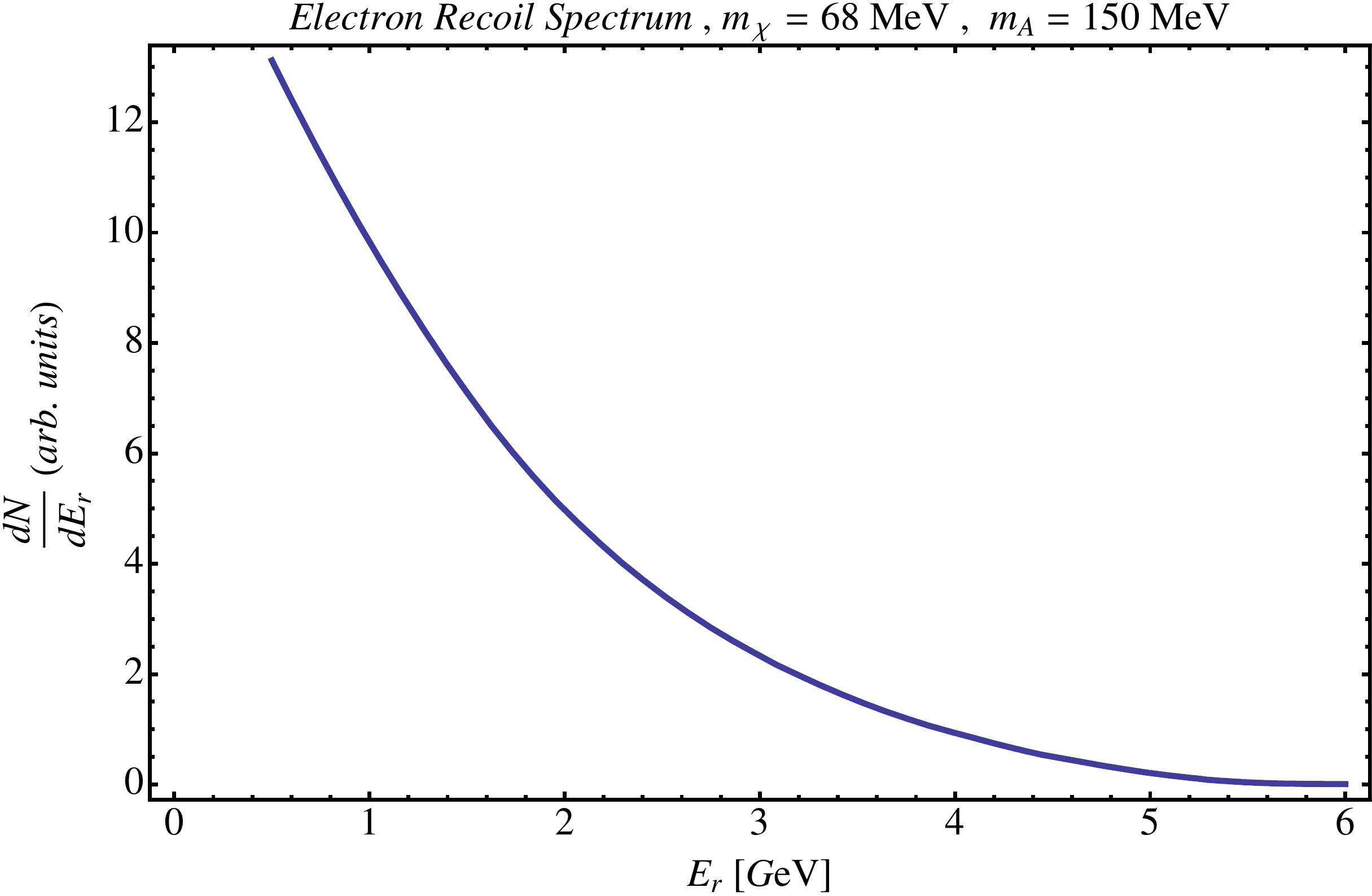}
 \caption{Energy distributions of scattered electrons for two choices of $m_\chi$ and $m_{A'}$. }
\label{fig:energyE} 
\end{figure}  
The characteristic momentum transfer in $\chi$-matter interactions is of order the $A'$ mass.
For $A'$ masses from tens to hundreds of MeV, $\chi$ can scatter coherently off a nucleus (with a $Z^2$-enhanced cross-section, but the lowest energy transfers and therefore highest radiological backrounds) or eject a nucleon through a quasi-elastic scattering reaction $\chi p,n \to \chi p,n$, with characteristic recoil energy $E_R \sim m_{A'}^2/{2 m_N} \sim 1-50$ MeV.  Resonant single-pion and non-resonant double-pion production processes may also have appreciable yields, particularly for higher $A'$ masses. 

A second signal which is slightly sub-dominant but of considerable experimental importance is $\chi$-electron scattering.  Because the electron is light and the $\chi$ are energetic, scattered electrons typically carry GeV-scale energy and are therefore subject to much lower backgrounds than nucleon scattering.  
Figure~\ref{fig:energyE} shows the lepton recoil energy for different choices of $m_\chi$ and M$_{A'}$.
Indeed, for models with kinetically mixed mediators which produce both electron- and nucleon-scattering signals, it is likely that electron-recoil searches at BDX will have the greatest sensitivity.  At the same time, other models appear to have distinctive signals primarily in nucleon-scattering, for which BDX's unique sensitivity to nuclear recoils is ideal.  

To close this discussion of dark matter models and their signals, we comment on a simple and well-motivated generalization of the model introduced in \eqref{eq:lagrangian}.  The dark matter particle $\chi$, which in \eqref{eq:lagrangian} was given a Dirac mass, can further acquire a Majorana mass --- for example, from a coupling to the $U(1)_D$ Higgs field --- as in \cite{TuckerSmith:2001hy}.  The Majorana mass term splits $\chi$ into two mass eigenstates ($\chi$ and $\psi$) with a mass splitting $\Delta$; this theory has an $A'-\chi-\psi$ interaction, but no mass-diagonal $A'$ couplings.  As a result, (a) the electron beam produces $\chi \psi$ pairs, (b) for sufficiently large mass splittings $\Delta$, the $\psi$ decays promptly to $\chi e^+ e^-$, and (c) the  $\chi$-scattering processes in the detector are inelastic (e.g. $\chi p \rightarrow \psi p$), with a total deposited energy that is often dominated by the energetic $e^+e^-$ pair from the subsequent $\psi$ decay. 
Like the electron scattering process, this inelastic scattering signal can be searched for with very low background rates.

\begin{figure}[t]
\center
\includegraphics[width=11.0cm]{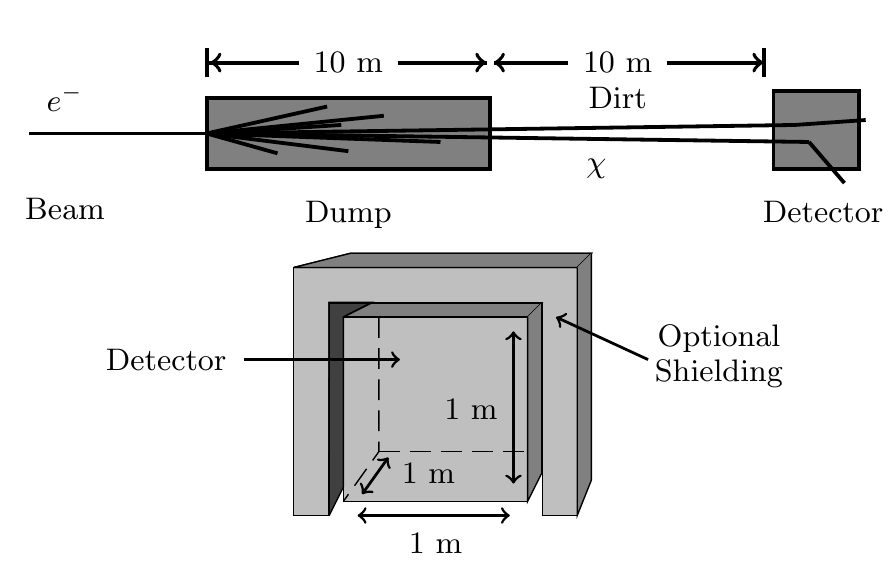}
\caption{Schematic of the experimental setup. A high-intensity multi-GeV electron beam impinging on a beam-dump produces a secondary beam of dark sector states.  In the basic setup, a small detector is placed downstream with respect to the beam-dump so that muons and energetic neutrons are entirely ranged out. 
 }\label{fig:setup}
\end{figure}

\subsection{Future beam-dump experiments}
Many experiments have been performed searching for an $A^{\prime}$ boson with mass in the range $1 - 1000$ MeV and coupling, $\varepsilon$, in the range $10^{-5} - 10^{-2}$.
Several different and complementary approaches were adopted (for a summary, see \cite{Bjorken:2009mm} and references therein),  but  no positive result was found so far. 
All exclusion limits are based on the assumption that dark-photons decay to $e^{+} e^{-}$ only or, alternatively, that they couple with the same strength to all fermions

If other $A^{\prime}$ decay modes are accessible, counts associated to visible decay would scale as $\varepsilon^{4} \alpha / \alpha_D$ rather than $\varepsilon^{2}$, and  the $A^{\prime}$ width would increase. Accounting for both effects, none of previous searches implies bounds on the coupling constant stronger than $\varepsilon^{2}\simeq 10^{-3}$ .
For similar reasons, the beam-dump limits on weakly coupled $A^{\prime}$, that rely on an $\varepsilon^{2}-$suppressed $\apr$ width, do not apply. Experiments searching for \textit{invisible} $A^{\prime}$ decays are therefore of particular interest being  complementary to the experimental program searching for   dark-photon: a full test of the dark sectors idea requires in fact searches sensitive to \textbf{all} possible $\apr$ decays, visible and invisible.
Moreover, important regions of the parameter space where an $A^{\prime}$ can explain the $g-2$ anomaly are still viable for $A^{\prime}$ invisible decay channels, and these can be explored by BDX. The improved sensitivity of BDX over other experiments will allow BDX to explore $g-2$ parameter space at smaller values of $\alpha_D$ than are currently probed (by E137 for example). 
Below, a list of experiments  that were recently proposed is reported.

The PADME \cite{padme} experiment proposes a search for invisible $A^{\prime}$ decays in the process $e^{+} e^{-} \rightarrow A^{\prime} \gamma$. The 510 MeV positrons from the DA$\Phi$NE LINAC would  scatter on atomic electrons of a thick target, and only the final state photon would be detected. 
The $A^{\prime}$ is thus reconstructed via the final state photon missing mass. The $A^{\prime}$ mass range that can be explored is limited to 22.8 MeV due to  the small energy available in the center-of-mass frame. Nevertheless, with $10^{13}- 10^{14}$ positrons on target per year, PADME wold set a limit down to $10^{-8} - 10^{-9}$ in $\varepsilon^{2}$, without making \textit{any} assumption on the $A^{\prime}$ decay, and thus exploring a completely unknown region of the parameters space. A similar experiment has been proposed at the VEPP-3 storage ring \cite{Wojtsekhowski:2012zq}, with 500 MeV positrons impinging on an gas hydrogen internal target,  detecting the final state photon only. In 6 months of run at a luminosity of  $10^{32}$ cm$^{-2}$ s$^{-1}$, the experiment can explore the parameter space region below $m_\apr \simeq 20$ MeV and down to $\varepsilon^{2} \simeq 10^{-8}$.

A recently proposed experiment at CERN SPS \cite{s2013lya} would also search for invisible $\apr$ decays. The experiment employs an innovative technique, by having the primarily $e^{-}$ beam from the SPS, with energy between 10 and 300 GeV, impinging on an \textit{active} beam-dump, made by a calorimeter based on scintillating fibers and tungsten, ECAL1. A nearly-hermetic detector would be  located behind the active beam-dump. The detector is made by a charged particle veto counter, a decay volume, two scintillating fiber counters, a second electromagnetic calorimeter ECAL2, and an hadronic calorimeter. The primary goal of the experiment is to search for $\apr$ production in the active dump trough a Bremmstrahlung-like process, followed by $\apr$ decay to $e^{+} e^{-}$. The signature for these events is a signal in ECAL1 and two clusters in ECAL2, from the $\apr$ decay products. 
The same experiment, could  also  search for $\apr$ invisible decays by exploiting the detector hermeticity, and requiring a single hit in ECAL1 from the  $e^{-}$ radiating  the $\apr$. The projected sensitivity for $3\cdot 10^{12}$ electrons on target covers a very large region in the $\apr$ parameter space, with $m_{\apr}<1$ GeV and $\varepsilon>1\cdot10^{-5}$. However, in the event of a positive signal, the experiment would carve out a contour in the parameter space, 
but would not independently measure the $\apr$ properties,

Direct searches for $\apr$ decay products are natural extensions of  the $\apr$ invisible decay search program presented above. 
Experiments measuring the $\chi$ scattering in a detector placed at tens of meters downstream of the dump of a high-intensity, high-energy beam have access to the full $\apr - \chi$ parameter space,
and have the potential of constraining a vast part of it.
\begin{figure}[t!] 
\center
\includegraphics[width=12cm]{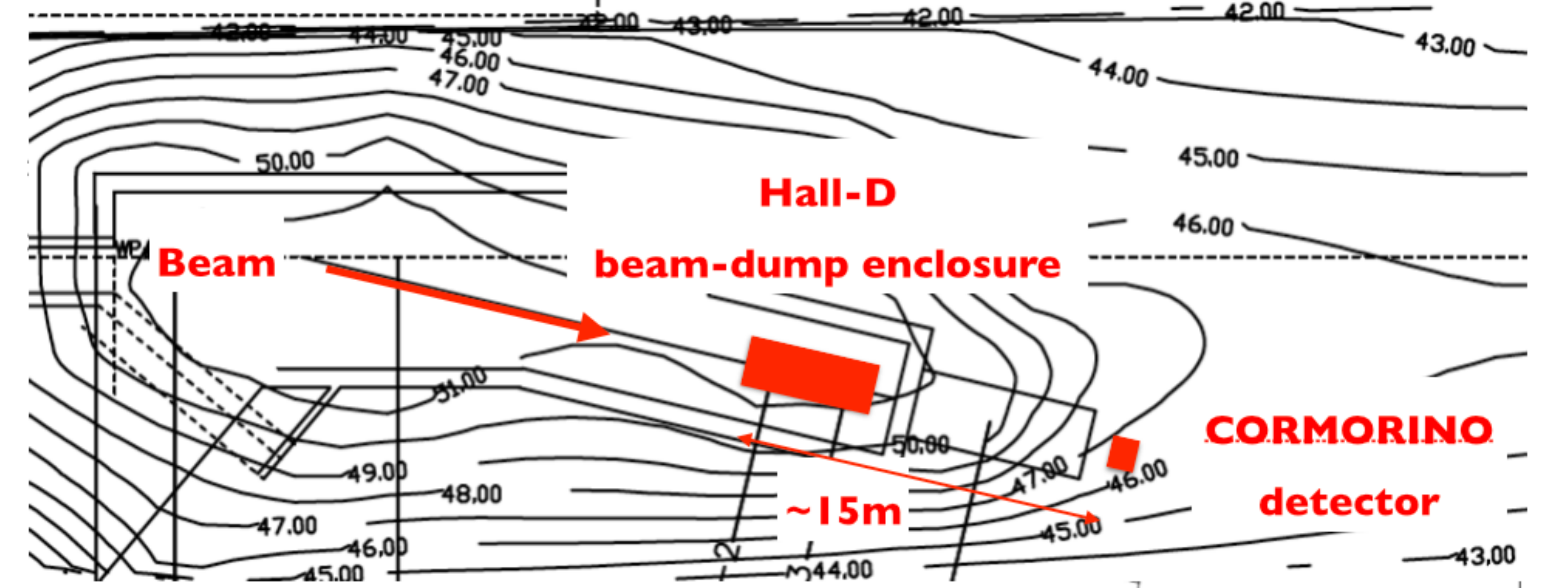}
 \caption{In the study case  the CORMORINO detector is deployed  $\sim$15m downstream the Hall-D beam-dump enclosure. The detector is placed above the ground within an iron shielding pit.}
 \label{fig:exp_setup}
\end{figure} 
The MiniBooNE experiment, originally designed to study neutrino oscillations, recently completed a test run to demonstrate the feasibility of an MeV-scale $\chi$s \cite{MiniBoone2013} search.
During the test run, the primary 8.9 GeV proton beam from the FNAL accelerator impinged on a 50-m long iron beam-dump. 
The dark matter particles are produced through neutral meson decays ($\pi^{0}$, $\eta$), in which one of the final state photons converts  to an $\apr$, which then decays to a $\chi \overline{\chi}$ pair. 
 These particles can then scatter on the electrons or nuclei in the MiniBooNE detector, placed 490 m downstream of the beam-dump. The otherwise dominant neutrino background,  generated by charged meson decays in flight, was reduced by a factor of $\simeq 70$ by diverting the proton beam into the dump, away from the original beryllium target. 
With the support of the FNAL PAC, MiniBoone has collected 1.1 $10^{20}$ protons on target in this beam-dump mode between October 2013 and April 2014.  They will continue to take data in this configuration until September of 2014, and the team is evaluating analysis methods using the first two months of data.  

\subsection{A beam-dump experiment  at JLab}
The experiment proposed in this Letter of Intent would require a $1\ \m^3$-scale (or smaller) detector volume, located at tens of meters downstream of the dump of a high-intensity 
multi-GeV electron beam, and could run parasitically (see \cite{Diamond:2013oda} for a proof-of-concept example).  
See Fig.~\ref{fig:setup} for a schematic representation of the experimental setup.
The experiment will use {\it both} electron scattering and low-energy nuclear recoil signatures and exploit excellent forward geometric acceptance 
to greatly extend dark matter sensitivity beyond the parameter space covered by the high threshold/low acceptance E137 setup or by existing proton beam-dumps. 
The approach also takes advantage of Jefferson Lab's upgrade  to 12 GeV energies with the new CEBAF, which is 
scheduled to deliver up to  $\sim 100 \mu$A currents. 

\begin{figure}[t!] 
\center
\includegraphics[width=12cm]{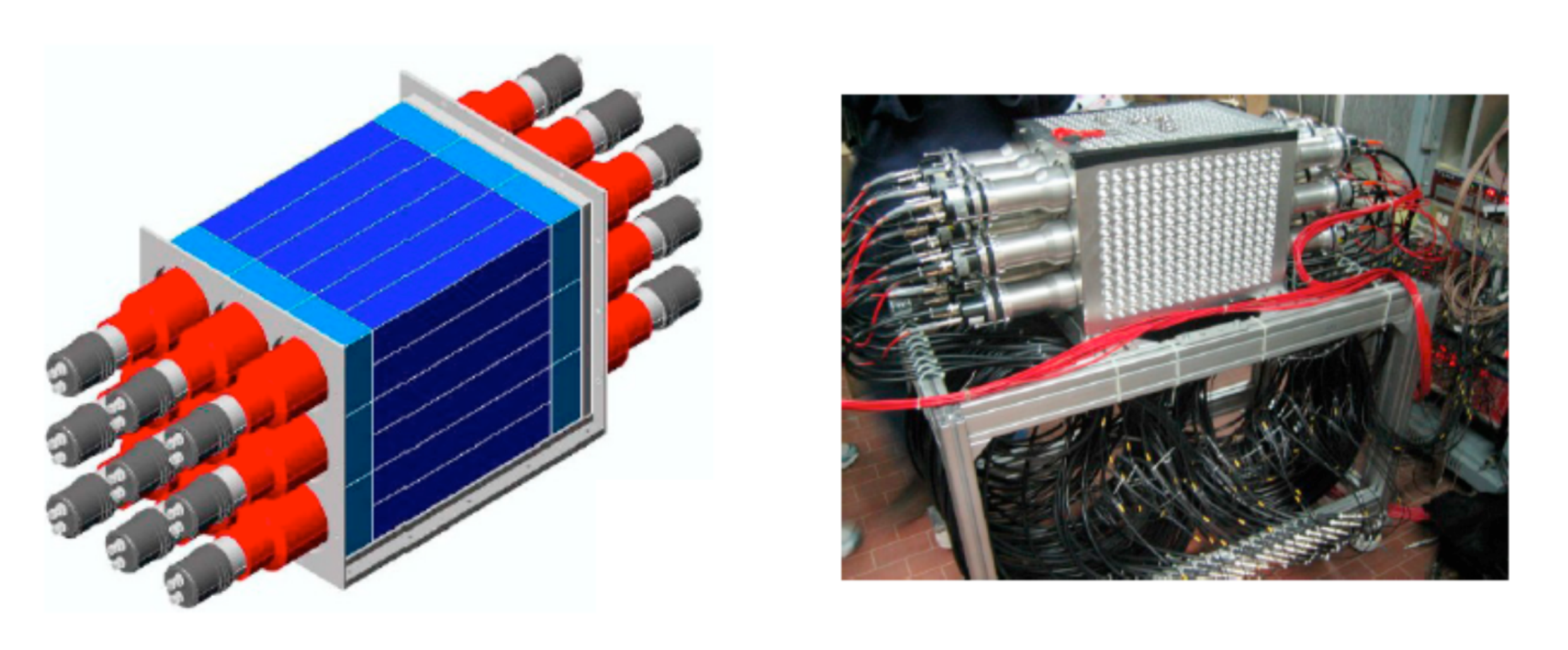}
 \caption{ (Left) CAD representation of the CORMORINO detector, made by 9 bars of plastic scintillators read by 3PMTs, each read on both sides by a a 3'' PMTs . (Right) A picture of the detector mounted on its support and cabled.
 }\label{fig:cormorino}
\end{figure}

\begin{figure}[t!] 
\center
\includegraphics[width=14cm]{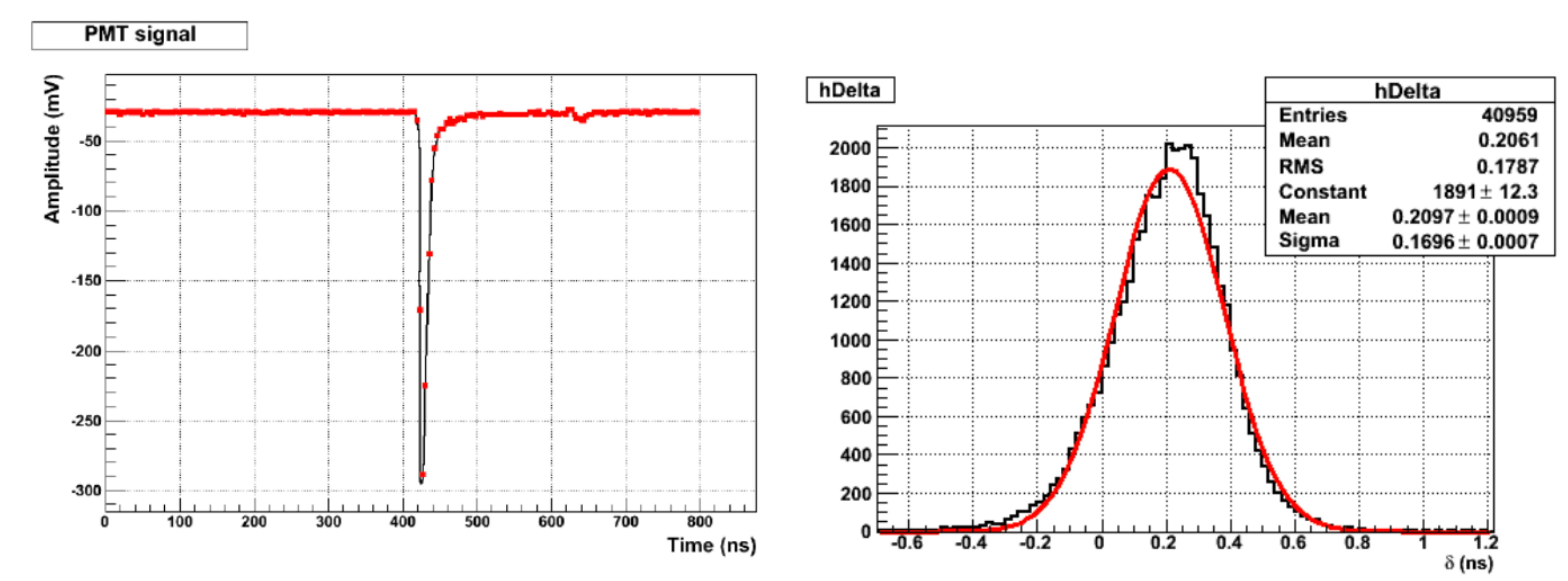} 
 \caption{Left: a typical signal from CORMORINO and its digitization. Right: time resolution of CORMORINO measured with cosmic muons. }\label{fig:signal}
\end{figure}
\section{A study case: CORMORINO and JLab Hall-D beam-dump}  
In this Section we present the results of a study for a possible beam-dump experiment at JLab.
This case study places an existing detector, CORMORINO, downstream of the Hall-D beam-dump, as shown in 
Fig.~\ref{fig:exp_setup}. 
We study this option in detail because CORMORINO  implements a technology that we consider as a viable option for a full scale experiment 
and that,  with  minor modifications, can be easily used to measure cosmogenic backgrounds and validate simulations.
Beam-related backgrounds were simulated in the Hall-D beam-dump configuration because, being above ground, this represents
 a set-up where  CORMORINO could be easily installed  with no need for major civil construction to perform a test run and validate the MC estimates. 
Based on these considerations, in this Section we report estimates for beam-related and beam-unrelated background rates in this experimental configuration as well as the expected signal rates. 
These results will be used to estimate the counts expected in a full experiment whose 
concept,  including a $\sim $1 m$^3$ optimized detector placed behind a high intensity Hall beam-dump (A or C), is  discussed  in the next section.

\begin{figure}[t!] 
\center
\includegraphics[width=14cm]{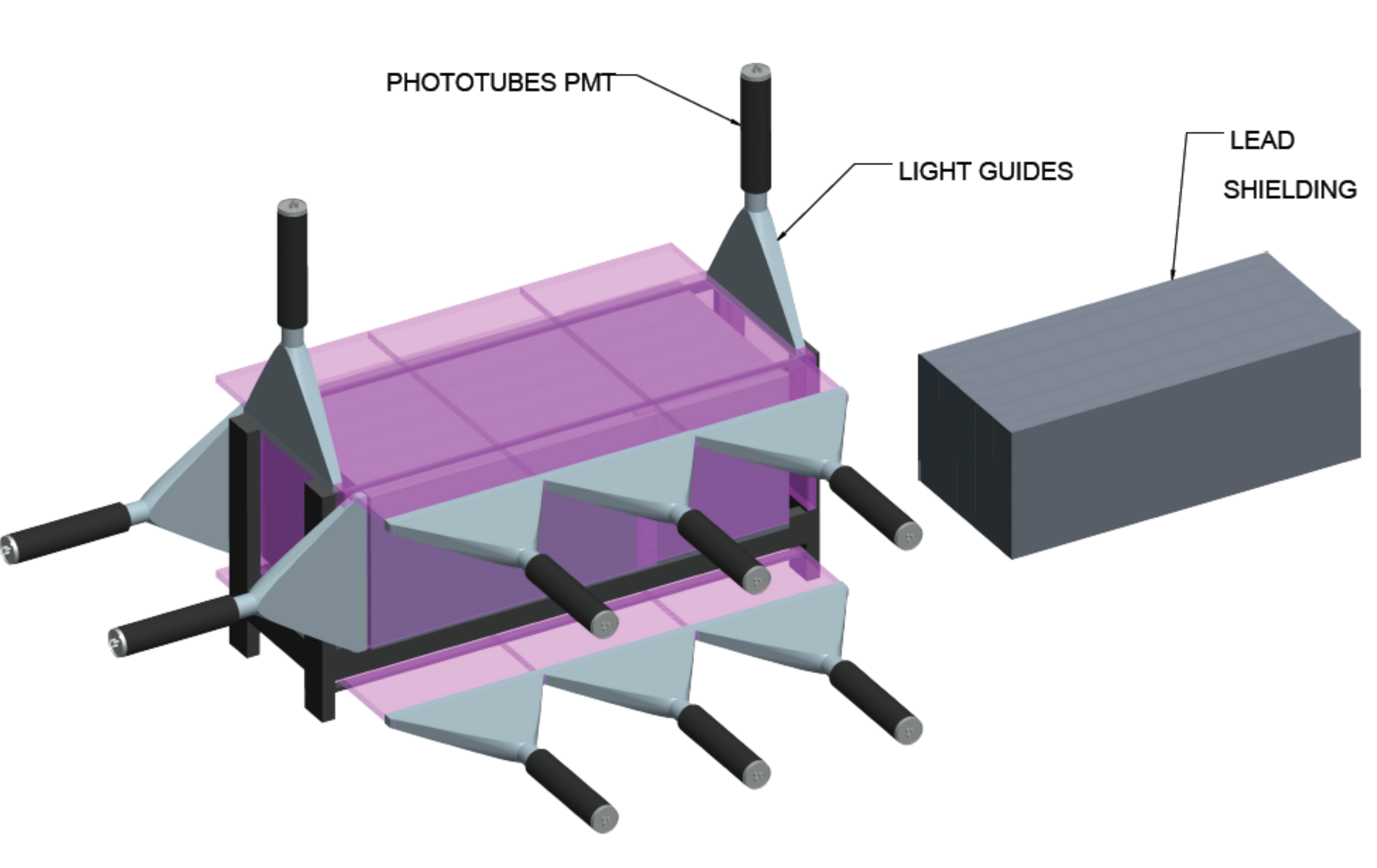} 
 \caption{The veto detector sourrounding CORMORINO is made by sheets of 2 cm thick plastic scintillators coupled to PMTs by shaped Plexiglas light guides. 
 A thin layer of lead shields gamma's produced by particles decaying between the veto and CORMORINO. This detector is currently being assembled.} \label{fig:cormorino_veto}
\end{figure}

\subsection{The CORMORINO detector}
The CORMORINO detector is a 0.036 m$^3$ cube made of plastic scintillator bars, whose output is read by PMTs. 
It is a prototype built for the experiment CORMORAD (COre Reactor MOnitoRing by an Antineutrino Detector), which aims to use antineutrino to monitor the power and isotopic composition of the core of a commercial nuclear power plant. The 1.5 - 10 MeV antineutrinos  produced by the fragment of the  nuclear fission interact with protons of the plastic scintillator by inverse beta decay, thereby producing a positron and a neutron. 
The plastic scintillator, acting as an active target, provides a high number of protons and good detection capability for both positron ionization and neutron capture. 
The detector contains nine optical channels, each consisting  of four 5x5x30 cm$^3$ plastic scintillator bars  (NE102) individually wrapped in aluminized Mylar.
  The four bars are coupled on each side to a  Photonis XP2312 3''  photomultiplier via a 10x10x5 cm$^3$ scintillator block,  acting as light guide. Figure~\ref{fig:cormorino} shows a CAD drawing and a picture of the detector. The 9 (Left) +  9 (Right) PMTs are connected to 250 MHz 16ch JLab fADC boards for signal digitization.
A typical signal is shown  on the left panel  of Fig.~\ref{fig:signal}. In the same Figure on the right,   the time resolution measured with cosmic muons is shown. With the current PMT gain, an energy deposition in the plastic scintillator  of 1 MeV corresponds to a signal of 8 mV in amplitude in each PMT. Despite the aged plastic scintillator used in CORMORINO, the measured time resolution for MIPs is still $\sigma_T \sim 110$ ps.
With new plastic, better optical coupling and optimized PMTs a time resolution of $\sim 50$ ps  can be easily obtained as demonstrated by the new CTOF detectors built for CLAS12~\cite{clas12-ctof}. 
The low rate expected in the experiment does not require a selective trigger and therefore the DAQ scheme that is  implemented
uses a simple L$\&$R single-bar coincidence to save the full wave shape stored in the fADC. The trigger threshold can be set at 5 mV (corresponding
 to the energy deposited by a  600 keV electron),which leaves more sophisticated selection criteria for the off-line analysis to reduce backgrounds \footnote{In fact, having the full shape for each sample it is possible to do some particle identification, time measurement and time correlation between samples to reject neutron capture, rare muon decays, etc. }. We are currently building a hermetic charged-particle veto counter based on 2 cm thick plastic scintillator planes to be placed  in front of each of the 6 faces of the CORMORINO to reject cosmic muons and further reduce backgrounds. A layer of 5 cm of lead between the veto and the detector surface will shield from low energy gamma
 from particles decaying inside the external iron shield. Figure~\ref{fig:cormorino_veto}  shows the layout of the active and passive veto surrounding CORMORINO.

\subsection{The Hall-D beam-dump}
The Hall-D at JLab is expected to receive from CEBAF a 12 GeV electron beam with an average current of  $\sim$200 nA.
This moderate current  is not  the optimal choice for a full scale experiment at JLab, but the simplified logistics motivated us to study this configuration in the event that beam-related backgrounds 
were found to be dominant, calling for a test run. The beam line delivering electrons to Hall-D runs above the ground in the last section of its path, in proximity of the beam-dump, which simplifies any possible detector deployments behind the beam-dump enclosure with access from all sides. To set up the experiment  a hut hosting the detector in a water-tight and air-conditioned environment with few services such as power and network connection is required. For this option a shield for cosmogenic neutrons and cosmic rays needs to be artificially implemented, providing a pit of few meters of iron all around the detector. 

\begin{figure}[t!] 
\center
\includegraphics[width=16.5cm]{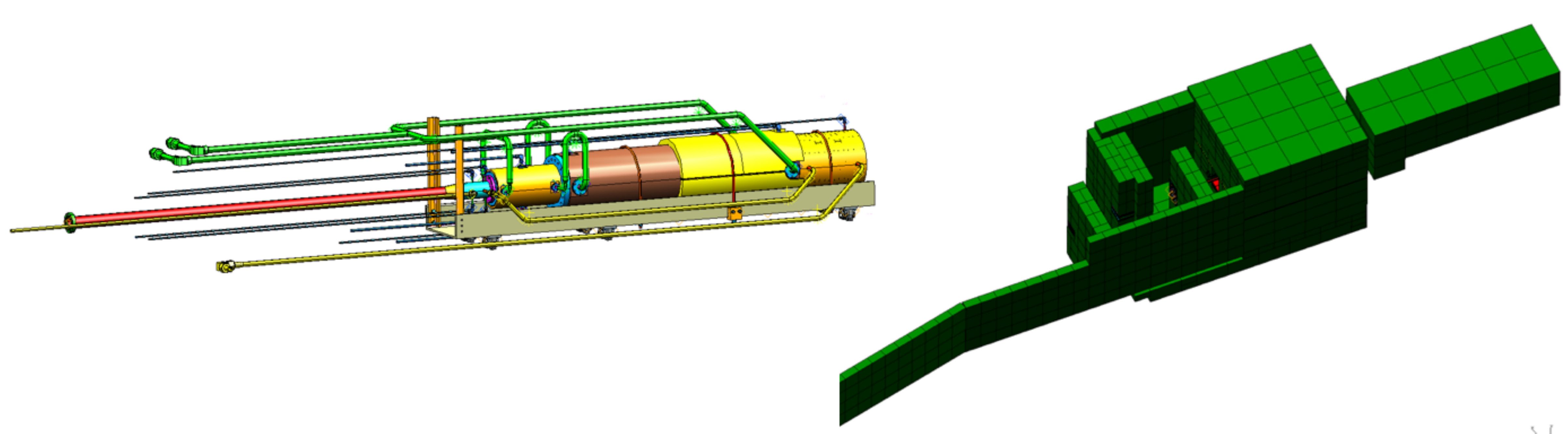}  
 \caption{ (Left) The Hall-D beam-dump;  the aluminum and copper cylinder where the beam stops and the cooling pipes surrounding the dump are clearly visible. (Right) The beam-dump enclosure:  only the external concrete wall is visible while the iron shield is hidden inside. }\label{fig:halld_bd}
\end{figure}

The Hall-D beam-dump consists of an aluminum cylinder, $\sim$25 cm in diameter and $\sim$100 cm in length, followed by a  $\sim$30 cm long copper cylinder of the same diameter. They are both cooled by circulating water. The full drawing of the beam-dump is shown in the left panel of  Fig.~\ref{fig:halld_bd}. The beam-dump is housed in a concrete ($\sim$0.8 m thick) and iron ($\sim$4 m thick) enclosure shown in the right panel of the same figure.
To further reduce  possible radiations, another shield made by  $\sim$8 m of iron has been added downstream.

\subsection{Simulations of the  experimental set-up}\label{sec:expsetup}
\begin{figure}[t!] 
\center
\includegraphics[width=0.4\textwidth]{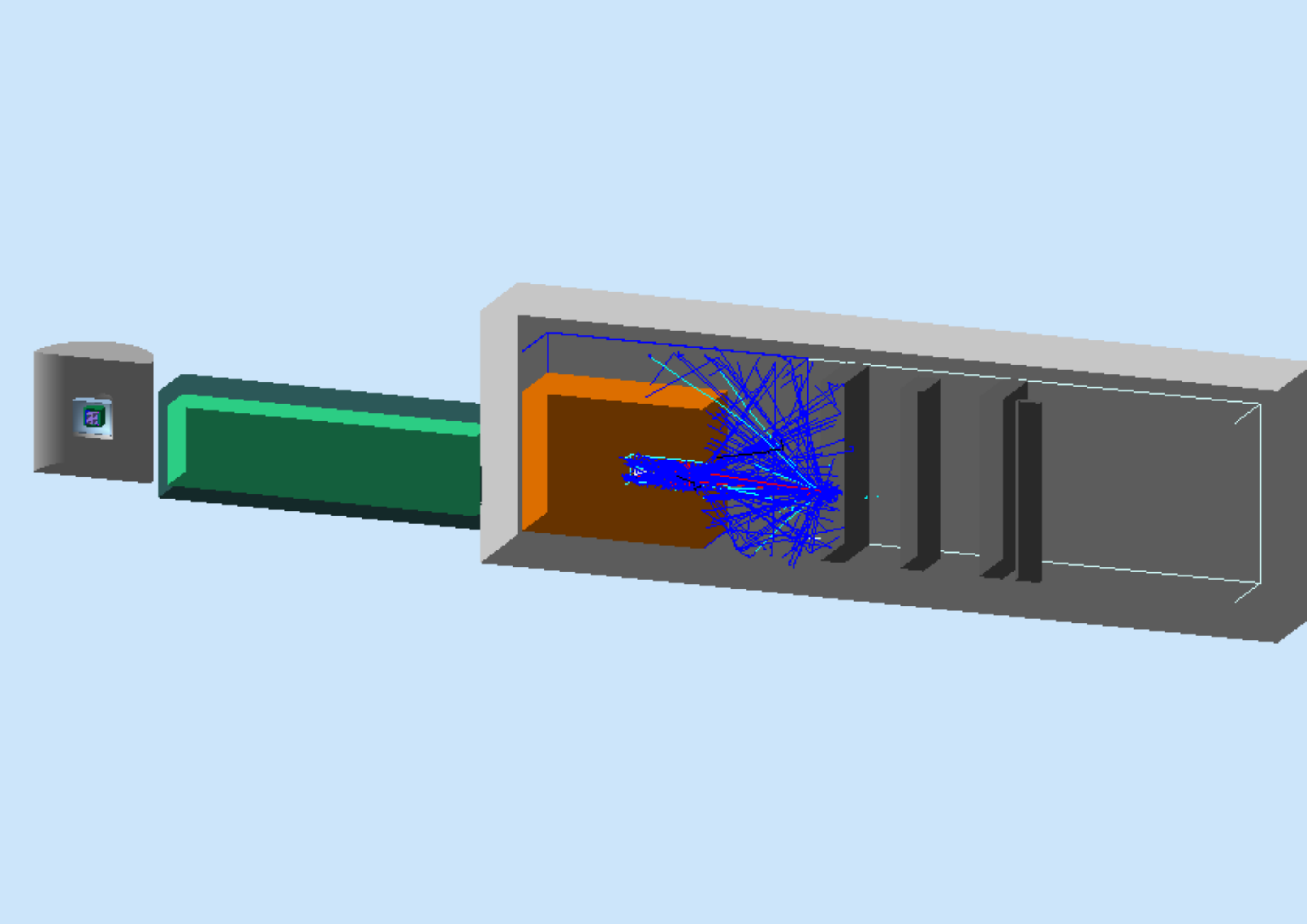} 
\includegraphics[width=0.285\textwidth]{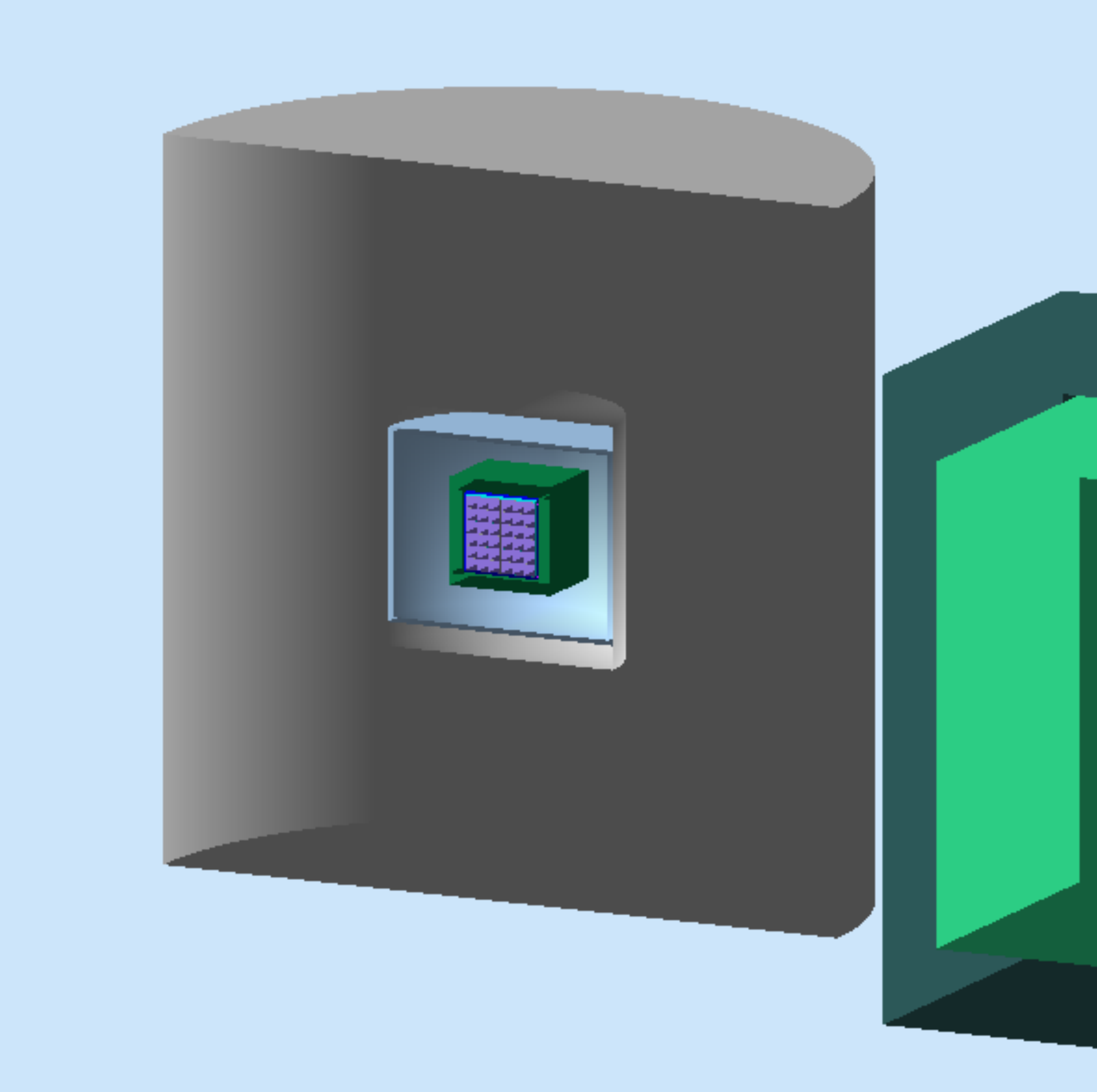}
\includegraphics[width=0.289\textwidth]{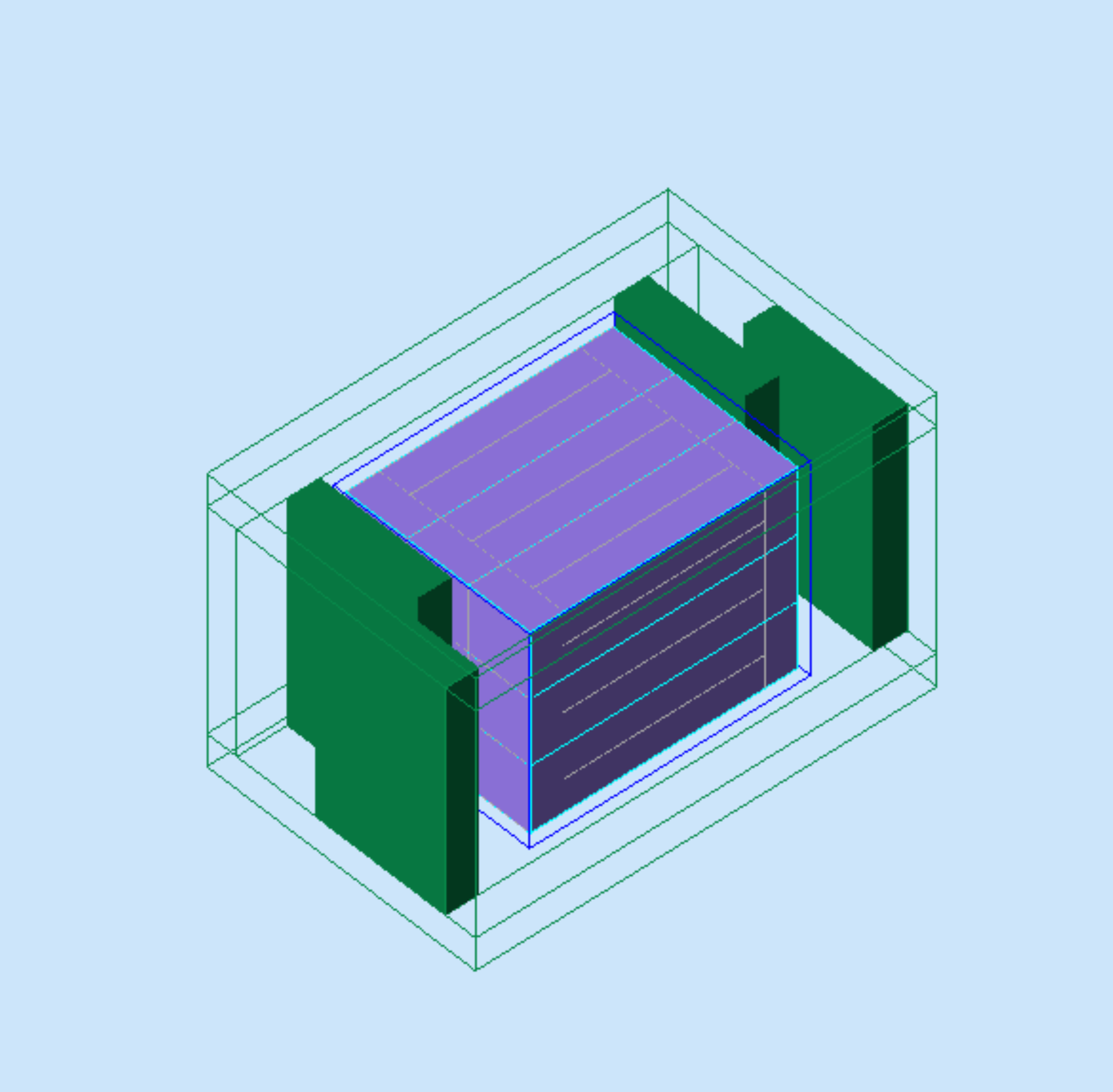}

 \caption{Implementation of the CORMORINO detector in GEMC. The section on the right shows the 9 optical channels (each made by 4 plastic scintillator bars), with a cylinder surrounding 
 the detector acting as an active veto. In the middle plot, the lead shielding (in green) is shown. The  section on the left shows the detector and veto inside a 1 m thick iron pit to reduce cosmogenic background.   }\label{fig:cormorino_gemc} 
\end{figure}

Both the CORMORINO detector and the Hall-D beam-dump geometry have been implemented in GEANT4 within the GEMC simulation package~\cite{gemc}. In the following sections we 
present some results concerning the beam-related background, the cosmogenic background, and the expected rates from an elastic interaction of a $\chi$ particle.  
The $\chi$ signal was simulated assuming the dominance of elastic DM interaction as discussed in Sec.~\ref{sec:elastic}.
Figure~\ref{fig:cormorino_gemc} shows the detector as implemented in the simulations. As mentioned above,
CORMORINO is positioned $\sim$15 m downstream the beam-dump, surrounded by a 1 meter of iron acting as a passive shield for cosmogenic neutrons and gammas. 

\begin{table}[htdp]
\caption{Relevant parameters of the theory for the two scenarios. The resulting number of $\chi$ produced per electron on target  (EOT) and the elastic interaction cross section $\sigma_{\chi-p}$ are also reported.}
\begin{center}
\begin{tabular}{|c|c|c|}
\hline
 & S.I &  S.II\\
\hline\hline
 $m_\chi$ & 10 MeV &  68 MeV\\
 \hline
$m_{A^\prime}$ & 50 MeV  & 150 MeV\\
\hline
$\epsilon$ & 10$^{-3}$ &10$^{-3}$\\
\hline
$\alpha_{D}$ & 0.1  & 0.1\\ 
\hline\hline
$N_\chi$ pairs produced per EOT& 3.4 10$^{-10}$ &3.4 10$^{-11}$  \\
\hline
$\sigma_{\chi-p}$ & 1.4 nb  &  0.14 nb\\
\hline\hline
\end{tabular}
\end{center}
\label{tab:twoscen}
\end{table}%

\subsubsection{$\chi$ production and  $\chi$-p elastic scattering in CORMORINO}
\begin{figure}[t!] 
\center
\includegraphics[width=16cm]{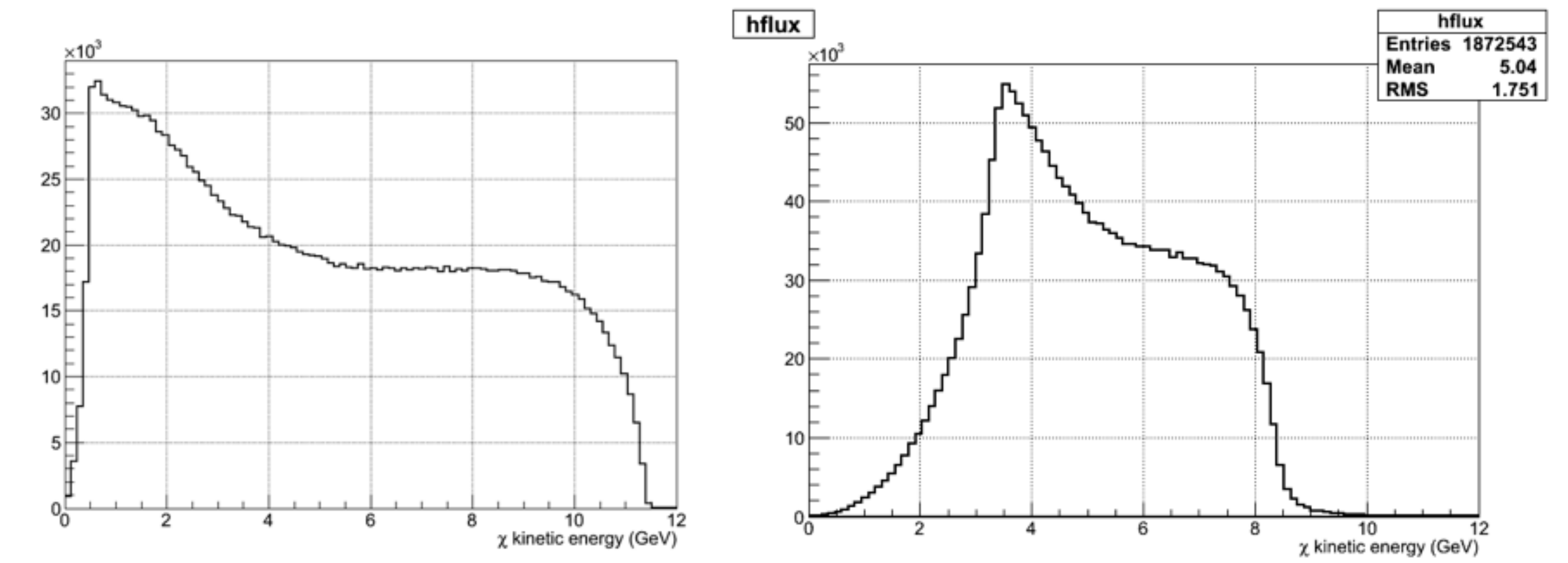}  
\includegraphics[width=16cm]{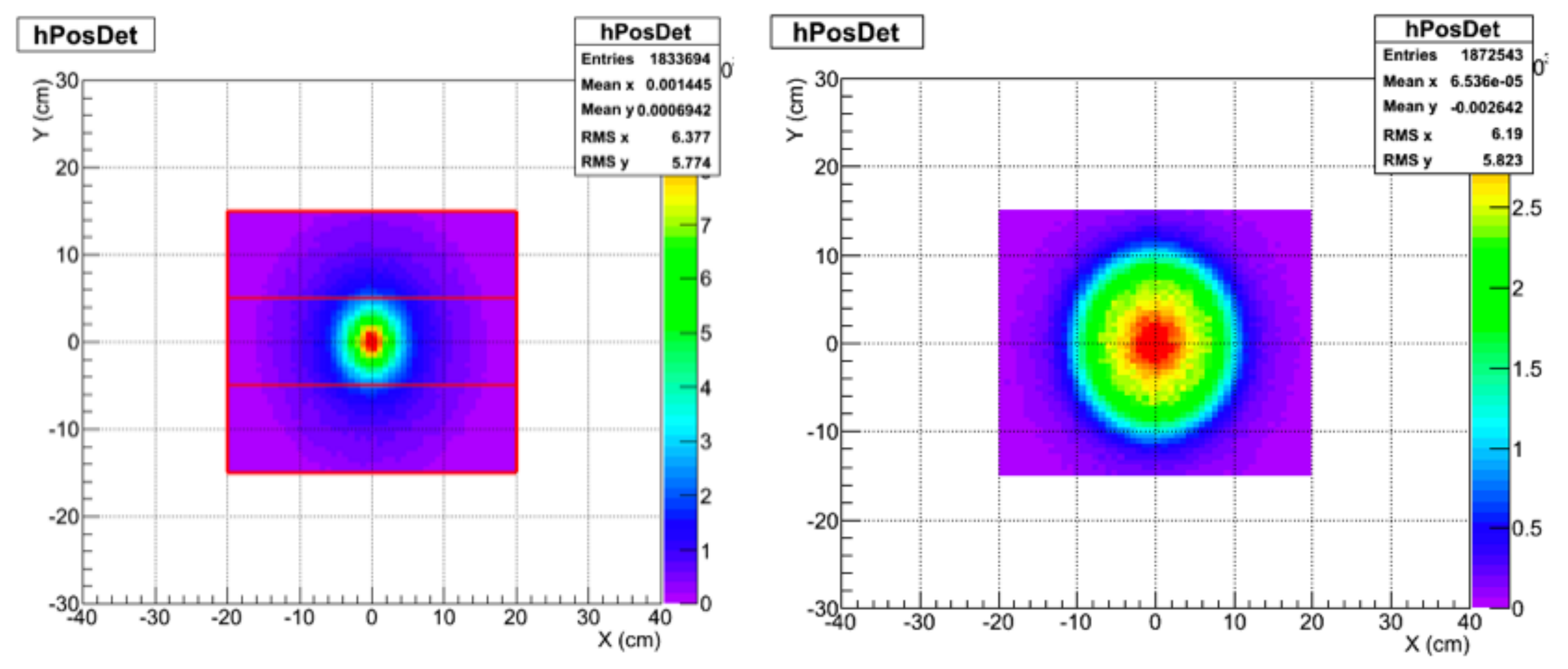}  
\caption{Top: the kinetic energy of  the $\chi$ particle  produced by 12 GeV electron beam in the two scenarios studied (left S.I, right S.II). 
Bottom: the distribution of $\chi$ events hitting the CORMORINO surface (left S.I, right S.II).}
\label{fig:kinea}
\end{figure}

\begin{figure}[t!] 
\center
\includegraphics[width=16.5cm]{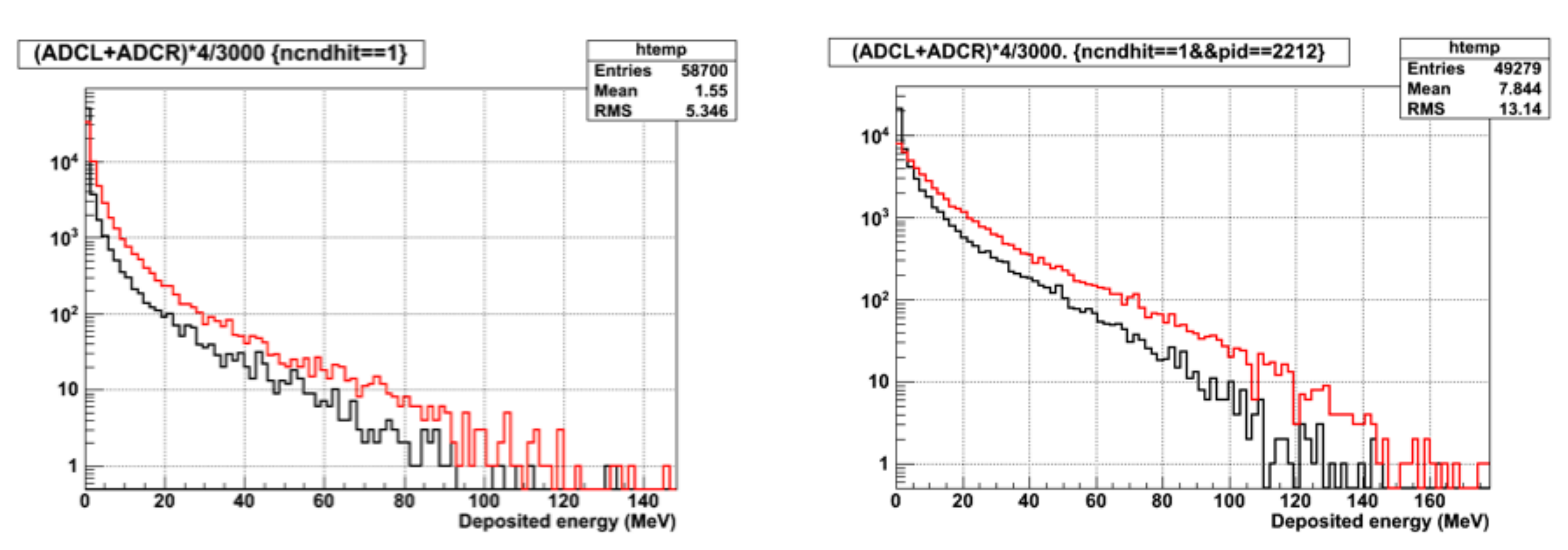} 
 \caption{The energy deposited (red) and reconstructed (black) in CORMORINO. The reconstructed energy spectrum includes the effect of light quenching in the two scenarios (left S.I, right S.II). }\label{fig:kinec}
\end{figure}

The detection of a $\chi$ particle in CORMORINO involves two steps: the $A^\prime$ electro-production and subsequent decay $A^\prime \to \chi \bar\chi$, occurring in the very first layer of the beam-dump,  and the elastic scattering of a $\chi$ on a proton of the detector.   Both processes depend on four
parameters: the mass of the $\chi$ ($m_\chi$), the mass of the exchanged $A^\prime$ ($m_{A^\prime}$), the coupling constant between the electron and the $A^\prime$
($\epsilon$) and the coupling constant between the $\chi$ and the $A^\prime$ ($\alpha_{D}$).  
We explored two possible scenarios, S.I and S.II, for which the corresponding parameters are  reported in Tab.~\ref{tab:twoscen}. The table also reports the resulting $\chi$ production rate (per EOT or {\it Electron On Target}) and the corresponding elastic cross section on a proton in the detector.
Figure~\ref{fig:kinea} shows the kinematics of the $\chi$ within the CORMORINO acceptance, assuming a 12 GeV electron beam (left and right plots correspond to S.I and S.II respectively). 
Figure~\ref{fig:kinec} shows the energy deposited in the detector by the  proton recoils  induced by the $\chi$-p elastic scattering. 
The red histogram shows the deposited energy while the black shows the reconstructed energy for the proton detection, including  the light quenching effect in the scintillator.
It is worth noting that a sizable fraction of events deposit a visible energy greater than 1 MeV.
To evaluate the number of  detected events in CORMORINO, we required a single hit defined as the coincidence of the PMTs coupled to the same scintillation bar.
For both scenarios we evaluate the rates for two thresholds: 1 MeV and 10 MeV\footnote{To realistically simulate threshold effects, we applied a threshold to each PMT, independently.}.
The effect of light attenuation in the scintillator was also included in the simulations, even if the effect is known to be small give the size of the detector.
The expected $\chi$-p elastic detection rate (per $\mu$A)   for a  threshold of 1 MeV (10 MeV) is therefore:
\be
R_{elastic \;\chi}^{S.I}=   1.0\; 10^{-5} {\rm Hz}/ \mu {\rm A} \;\; (1.2 \;10^{-6} {\rm Hz}/\mu {\rm A})
\ee
\be
R_{elastic \;\chi}^{S.II}=   2.0\; 10^{-7} {\rm Hz}/\mu {\rm A} \;\; (0.7 \;10^{-7} {\rm Hz}/\mu {\rm A})
\ee

\subsubsection{Beam-related background}
\begin{figure}[t!] 
\center
\includegraphics[width=17cm]{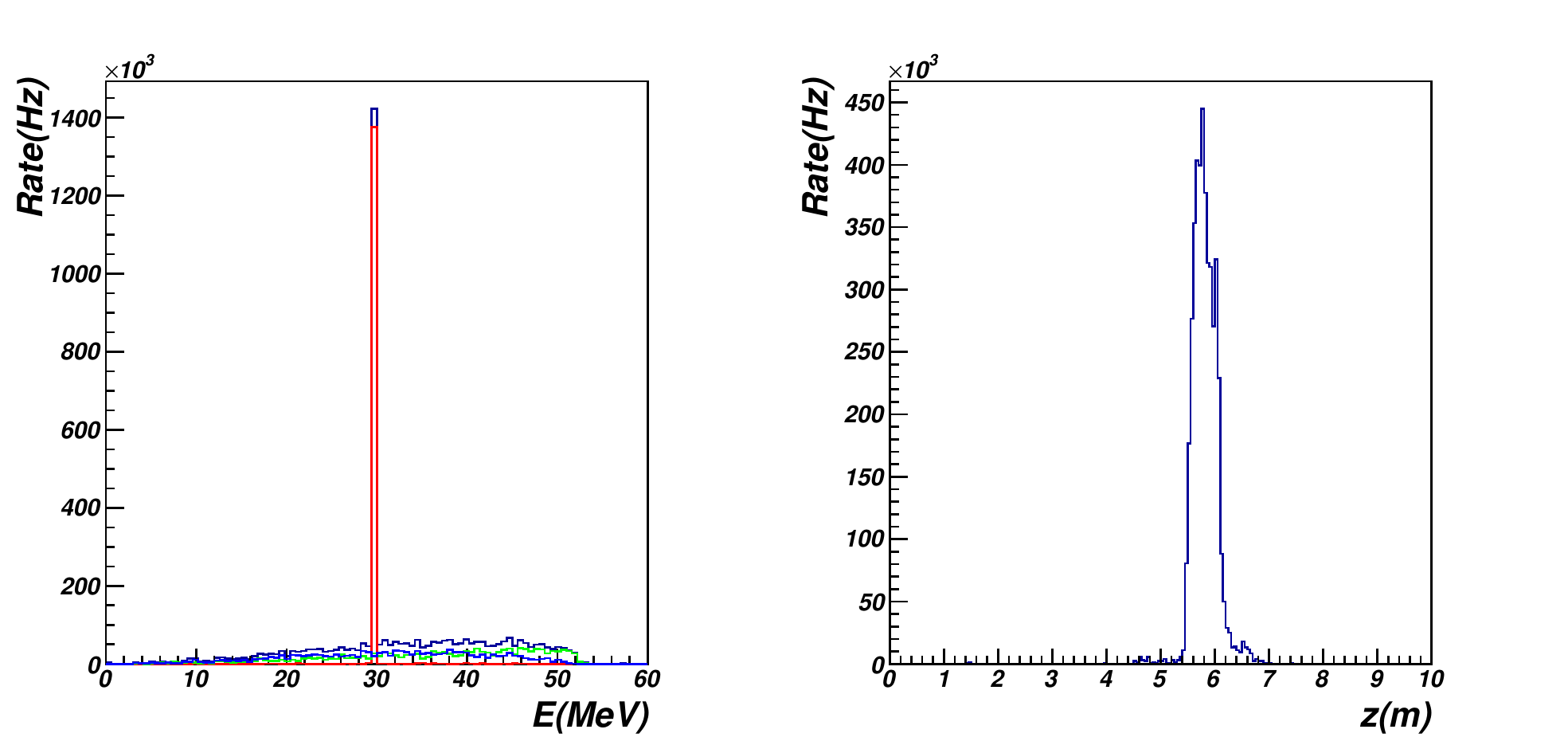}  
 \caption{Left: the energy spectrum of beam-related neutrinos crossing CORMORINO surface  (black: total spectrum; red: $\nu_\mu$, green: $\bar \nu_\mu$,  blue: $\nu_e$). Right: the vertex of the decaying primary particles ($\mu$ and $\pi$) starting at the beam-dump location and extending for 1 m downstream. }\label{fig:E_nu}
\end{figure}

The beam-related background has been simulated generating 1.6 10$^9$ 12 GeV EOT impinging on the beam-dump, corresponding  to 1.2 ms of 200 nA beam.
The only particles exiting from the beam-dump and crossing the detector area are $\nu$
 and $\bar\nu$ from $\pi^+$ decays
 ($\pi^+ \to \mu^+ \nu_\mu$) and $\mu^+$ decays ($\mu^+ \to e^+ \bar\nu_\mu \nu_e$) 
with a ratio of about 1.1 $\nu_\mu$ : 1 $\nu_{\bar\mu}$ : 1.1 $\nu_e$. 
 Negative pions are captured by nuclei before they can decay and therefore do not contribute to the neutrino counting. The energy spectrum of the 3 species is reported on the left panel of 
Fig.~\ref{fig:E_nu}: the positive pion decays at rest providing a monochromatic set of $\nu_\mu$ of 33.9 MeV (M$_\pi$-M$_\mu$). Neutrinos from $\mu^+$ decay share a total energy of 105.1 MeV (M$_\mu$-M$_e$), which results in a upper limit for the energy of the  $\nu_{\bar\mu}$ and  $\nu_e$.  These neutrinos origins inside the beam-dump as shown in the right panel of Fig.~\ref{fig:E_nu}. 
The rates of neutrinos  crossing the detector area, approximately equal for the three species, are $\sim$7.5 10$^6$ Hz/$\mu$A while the integrated rate is  $\sim$2.2 10$^7$ Hz/$\mu$A.
Neutrinos with energy in the range 2-100 MeV mainly interact by CC interaction  ($\bar\nu p \to e^+$ n) with a cross section of about  $\sigma_{\bar\nu p}\sim10^{-40}$ cm$^2$.
Considering  the volume of CORMORINO and assuming a detection efficiency of $\epsilon_{e^+ detection}=50\%$ with 1 MeV threshold (5$\%$ for 10 MeV
threshold), the background rate   is given by:
\be
R_{B-rel}^{\nu} = R^{\nu} \;\sigma_{\bar\nu p}\; N_{Avogadro} 1/A \; \rho_{plastic} \; L_{CORMORINO} \; \epsilon_{e^+ detection}= 2\;10^{-9} {\rm Hz}/\mu  {\rm A} \;\;(2\;10^{-10}  {\rm Hz}/\mu {\rm A})
\ee
No neutrons or other particles are observed within the  generated statistics. This allows us to set only an upper limit on the corresponding rate that we will improve by increasing the Monte Carlo statistics. 

Beside the beam-related background estimated by MC simulations, in Sec.~\ref{sec:systematics} we discuss  additional  checks that could be performed during the run by changing the experimental set-up (e.g. off-axis and  on/off  measurement, noise on PMTs, .etc.) to determine systematic  effects that  we may have under-evaluated or neglected by MC simulations.

\subsubsection{Beam-unrelated backgrounds}
Beam-unrelated backgrounds are mainly due to cosmic neutrons, cosmic muons and their decay products, including  rare decays of muons producing gamma's between the passive shield and the active veto. 
In this Section we briefly present background reduction strategies that could be applicable to  a new, full-scale, optimized detector.
We also present some preliminary results of MC simulations specifically for the CORMORINO set-up. To validate  MC simulations and optimize the selection algorithms,
we are planning to perform an extensive campaign of measurements with the CORMORINO detector in the same experimental conditions of a 
real experiment (detector placed  above the ground, same shielding as reported here, etc).

\paragraph{Time-Correlation}
Beam-uncorrelated backgrounds can be rejected by requiring a time coincidence between the RF signal and the event recorded by the detector. However, even if the studied case of a plastic scintillator detector provides a sub-ns response to a $\chi$ scattering, the nearly-CW structure of the CEBAF beam weakens the effectiveness of this technique. In fact, the bunch separation expected for the 12 GeV implementation is $\Delta$T=4.0 ns to be compared with the measured time  resolution of CORMORAD of $\sigma_T\sim$200 ps for 1 MeV deposited energy. The background reduction factor, $R$, can be expressed as the ratio between the time coincidence window width (3$\sigma_T$)  and $\Delta$T providing for JLab-CORMORINO(-like) conditions:
\be
R=\frac{3 \sigma_T}{\Delta T}\sim 20\% ~.
\ee
It is worth noting that for  the conditions discussed in this paper with a detector placed 15 m away from the beam-dump, $\chi$ masses between 0 - 500 MeV result in a time-of-flight  spread below  $\sigma_T$=200 ps (0 - 170 MeV for $\sigma_T$=20ps). The same argument applies to the $\chi$ momentum spread: any  p$_\chi$ between 6 GeV and 100 MeV correspond to a time-of-flight spread below  $\sigma_T$=200ps (6 GeV -  350 MeV for $\sigma_T$=20ps). This
 suggests that,  for a fast detector ($\sigma_T\sim$70-100ps) the rejection factor for uncorrelated background can be reduced by up to a factor of ten. This will be further discussed in in Sec.~\ref{sec:r-and-d}. In what follows,
  we conservatively omit the rejection power of of timing cuts, so this background rejection factor needs to be added to the results for a full sensitivity projection.

\paragraph{Directionality} 
Beside time-correlation, directionality could help reduce the cosmogenic background. In fact the angular distributions of  cosmic muons 
(and their decay products)  and cosmic neutrons are peaked in the vertical direction transverse to the beam (the angular distributions are proportional to $\cos^2{\theta}$ and   $\cos^3{\theta}$ respectively) whereas 
the $\chi$s enter the detector preferentially in the beam direction. The simple detector described here (a plastic scintillator cube) 
does not provide directionality capability for single-hit interactions, but this capability should be considered in a full-scale optimized detector as reported in Sec~\ref{sec:r-and-d}.\\

In the following we report  rate estimates for all species of cosmogenic background as obtained by GEANT4 simulations.

\begin{figure}[t!] 
\center
\includegraphics[width=16cm]{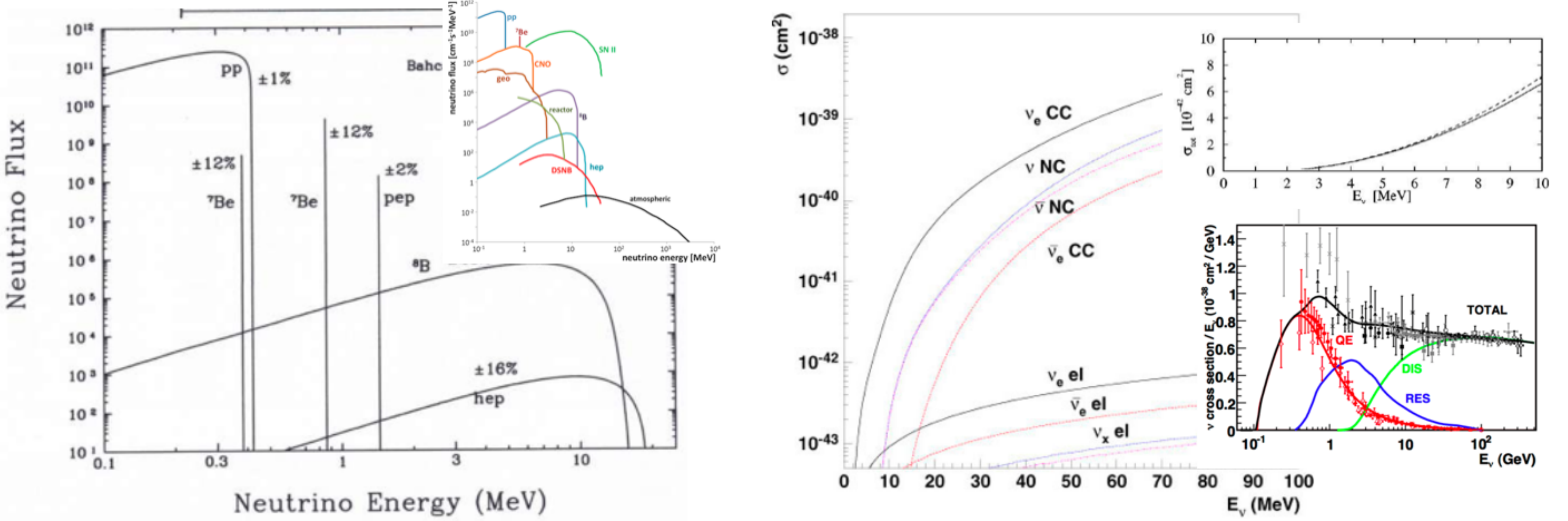} 
 \caption{Cosmic neutrino flux and interaction cross section. }\label{fig:neutrino}
\end{figure} 
\paragraph{Cosmic neutrinos}  
Cosmic neutrinos interactions represent a negligible contribution to the uncorrelated background. The only sizeable fluxwe need to consider is for solar neutrinos ~\cite{cosneutrino1} with energy below 10 MeV (see Fig.~\ref{fig:neutrino}-Left)  a rate on CORMORINO of $R^{Cosmic \nu}_{E_\nu<10 MeV}\sim 3 \;10^{11}$ Hz.  Low energy neutrinos (E$_\nu<$10 MeV)
 interact with protons in the  detector mainly by inducing inverse beta decay ($\bar\nu p \to e^+$ n), producing a positron that carries almost all the full
  neutrino energy, which could mimic a $\chi$ interaction. The corresponding cross section~\cite{cosneutrino2},  reported in Fig.~\ref{fig:neutrino}-Right) is in the range of  $\sigma_{\bar\nu p}\sim 3\; 10^{-42}$ cm$^2$.
The background rate,  considering  the volume of CORMORINO and assuming a detection efficiency of $\epsilon_{e^+ detection}=50\%$ 
 with 1 MeV thresholds (and 5$\%$ for 10 MeV threshold),  is given by:
\be
R_{Bg}^{\nu} = R^{Cosmic \nu}_{E_\nu<10 MeV} \;\sigma_{\bar\nu p}\; 1/A N_{Avogadro} \; \rho_{plastic} \; L_{CORMORINO} \; \epsilon_{e^+ detection}= 10^{-6}  {\rm Hz}\;\;  (10^{-7} {\rm Hz} )
\ee
The flux of high energy cosmic neutrinos  is several orders of magnitude smaller, which does not compensate for their larger interaction cross sections.
 Moreover, the elastic scattering  on a proton transfers only a small amount of energy, which reduces the probability of detecting the recoiling proton.
 For all these reasons the cosmic neutrino background is negligible compared to other sources.
\begin{figure}[t!] 
\center
\includegraphics[width=16cm]{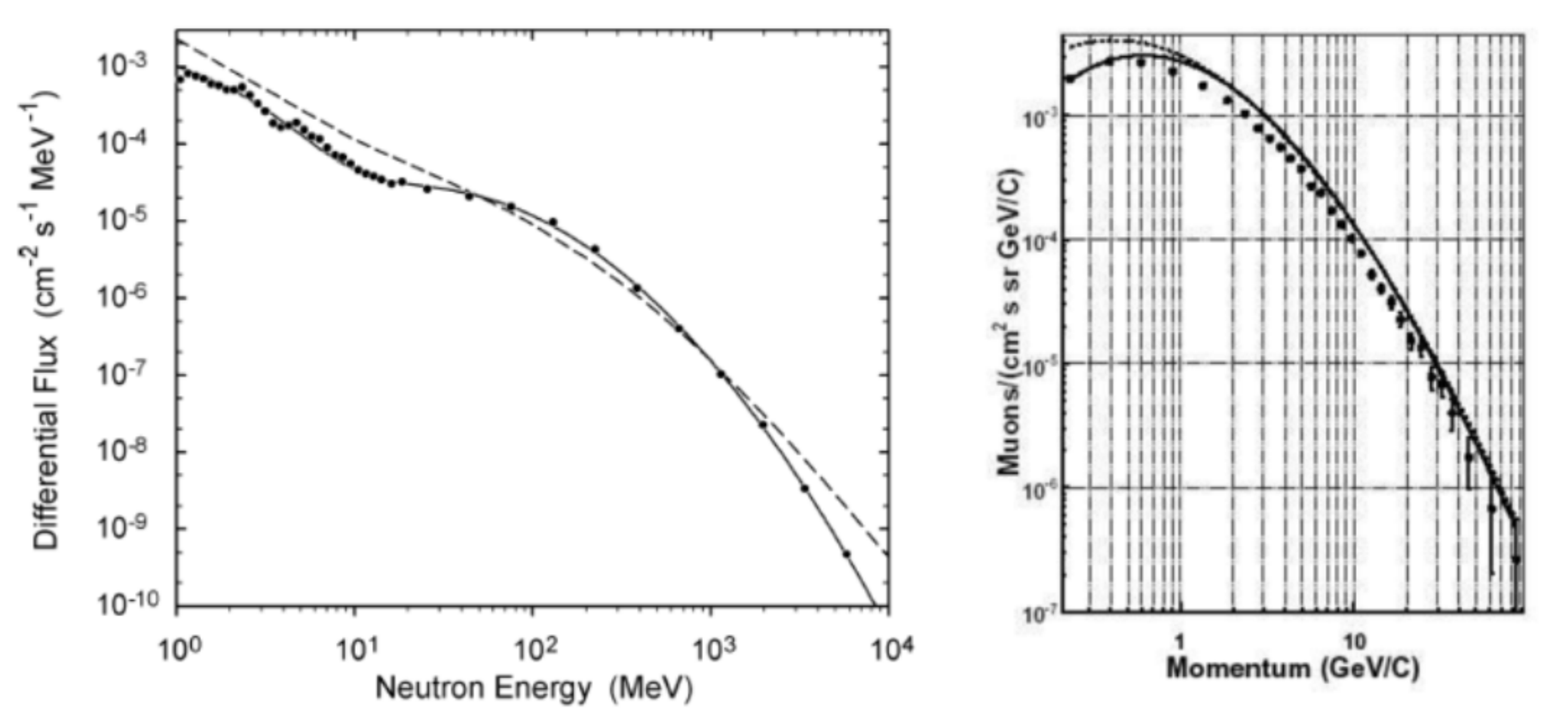} 
 \caption{Flux of cosmic neutron (left) and muons (right). }\label{fig:cosmic} 
\end{figure} 
 
\paragraph{Cosmic neutrons}
A cosmic neutron interacting with CORMORINO can produce a recoil above the trigger threshold, which looks identical to a recoil induced by  an incident $\chi$ particle. 
Considering the typical interaction length of neutrons in plastic, the 2 cm thick active veto has a small chance of detecting the incoming neutrons. On the other hand, the veto is 
sometimes useful to reject  neutron events, which produce hadronic showers  in  the heavy passive shielding when some charged fragments reach the detector.
 To estimate the rate of single hits in CORMORINO (mimicking a $\chi$ interaction), we generated a realistic
cosmic neutron flux as reported in Fig.~\ref{fig:cosmic}-Left with the  parametrization from Ref.~\cite{cosneutron}.
The overall neutron rate in open-air at  sea level (integrating over all energies and angles) is $\sim$64 Hz/m$^2$, (corresponding to a differential flux of  $\sim$50 Hz/m$^2$sr ).
The CORMORINO neutron detection efficiency $\epsilon_{n \; cosmic}$ depends on  the surrounding shielding, the incoming neutron energy ($E_n$) and  the detection threshold ($E_{Thr}$).  
Considering the 1 m  iron shield,  and a detection threshold of 1 MeV  (10 MeV) we found that  the detector is blind ($\epsilon_{n \; cosmic}\sim0$) to neutrons with energy lower than 50 MeV (100 MeV). 

Thus, the overall visible  rate for a detection threshold of 1 MeV (10 MeV) is:
\be
R_{Bg}^n=   2.7\; 10^{-3} {\rm Hz} \; \; (0.6\; 10^{-3} {\rm Hz} )
\ee

\paragraph{Cosmic muons}
Cosmic muons are by far the main source of beam-uncorrelated background  in a plastic scintillator detector. Fortunately, the most part of muons interact with both the veto counter and with CORMORINO, 
releasing the typical energy of a MIP, that can be tracked and rejected by exploiting the detector segmentation. Another source of background is due to muons decaying within the detector, between the shielding and the veto counter. In the following we report rate estimates based on MC simulations for  the measured muon flux shown in Fig.~\ref{fig:cosmic}-Right from Ref.~\cite{cosmuon}.

\subparagraph{Crossing muons}  
The combination of the active veto,  also sensitive to   cosmic muons crossing at large angles,  together with the CORMORINO identification capability (segmentation) sizably  reduces the number of crossing muons. 
The cosmic muon flux rate on the CORMORINO surfaces,  is 30.3 Hz (40$\%$  hits the top face and the remaining 60$\%$ hits the lateral sides) neglecting the  shielding.
The effect of 1 m iron shielding and 5 cm thick lead is to suppress   muons with momentum  $<$1.45 GeV/c, reducing the flux on the detector to R$^{All\;\mu}\sim$ 4.5 Hz, and 4 Hz (3.9 Hz) for hits   over a 1 MeV (10 MeV) threshold in CORMORINO.
Crossing muons usually induce 2 hits on the veto detector and  $\ge$1 hits in CORMORINO. 
To misidentify a crossing  $\mu$ as a $\chi$ it is necessary that  both hits in the veto counter are missed and only one CORMORINO bar records the hit.
Assuming the veto counter  inefficiency  to be (1-$\epsilon_{veto}$)=5$\%$  and  considering that $\alpha_{Single\; hit}^{Crossing \; \mu}$=32$\%$ for a 1 MeV threshold (36$\%$ for a 10 MeV threshold) of crossing muons release a visible energy in a single CORMORINO bar  (the rejection condition is N$_{Hits}^{Veto}$+N$_{Hits}^{CORMORINO}\ge$2), we end up with charged muon misidentification rate of:
\be
R_{Bg}^{Crossing \; \mu} = R^{All\;\mu}\; \alpha_{Single\; hit}^{Crossing \; \mu} \; (1-\epsilon_{veto})^2= 3.3\;10^{-3} {\rm Hz}  \;(3.5\; \;10^{-3} {\rm Hz} )
\ee
\subparagraph{Muons decaying inside CORMORINO}
Muons stopping inside the detector cross one side of the veto.  Positive muons stopping and decaying (at rest)  
 are  all detected for both  thresholds with rate R$^{\mu^+\;inside\;dec}\sim$ 0.12 Hz for  $\mu^+$. 
Negative muons, mainly captured by nuclei,  produce a visible signal in the detector  with rate R$^{\mu^-\;inside\;capture}\sim$ 0.12 Hz  for $\mu^-$   for both thresholds (1 MeV and 10 MeV).\\
Negative muons, if detected by the veto can be rejected. If not detected, and providing a single hit in CORMORINO, they contribute to the background. From MC simulations we derived
  $\alpha_{Single\; hit}^{\mu\;inside\;capture}$=24$\%$ for 1 MeV threshold and 39$\%$ for 10 MeV.
\be
R_{Bg}^{\mu^+\ \;inside \; capture} = R^{\mu^-\;inside\;capture}\; \alpha_{Single\; hit}^{\mu\;inside\;capture}\; (1-\epsilon_{veto}) =  1.4\; 10^{-3} {\rm Hz}\;\; (2.4\; 10^{-3} {\rm Hz} )
\ee
Positive muons that release energy above the trigger threshold (a fraction of R$^{\mu^+\;inside\;dec}$) will provide  a prompt signal (in coincidence with the veto) and a delayed hit 
due to the positron from $\mu^+ \to e^+ \bar\nu_\mu \nu_e$ . These events can be identified and rejected disregarding hits in the detector during a time window of
 10 $\tau_{Decay}=22 \mu$s after the prompt signal. This selection would  add  a negligible dead-time to the DAQ.
A source of background still arises from the veto inefficiency and the fraction of these events that provides a single hit (over threshold) in CORMORINO.
The resulting rate is:
\be
R_{Bg}^{\mu^- \;inside \; decay} = 2\; R^{\mu^-\;inside\;dec}\; \alpha_{Single\; hit}^{\mu\;inside\;dec}\; (1-\epsilon_{veto}) = 2.9\; 10^{-3}{\rm Hz} \;\; (4.8 \; 10^{-3} {\rm Hz} )
\ee
where the factor of 2 takes into account that the stopping muon and/or the positron can both trigger the detector (being $\alpha_{Single\; hit}^{\mu\;inside\;dec}=\alpha_{Single\; hit}^{\mu\;inside\;capture}$).

\subparagraph{Muons decaying inside lead shielding}
Here we consider muons that decay or get captured in the lead shielding between the veto and CORMORINO. We obtained a rate of  
R$^{\mu\;lead}\sim$ 0.9 Hz reduced to 0.3 Hz and 0.2 Hz for a detection  threshold of 1 MeV and 10 MeV respectively.
The fraction of events that deposited energy in a single CORMORINO bar 
is $ \alpha_{Single\; hit}^{\mu\;lead}\sim$0.46 and 0.43 for  a threshold of 1 MeV and 10 MeV respectively.
These values lead to a background rate of:
\be
R_{Bg}^{\mu^- \;lead} = R^{\mu\;lead}\; \alpha_{Single\; hit}^{\mu\;lead}\; (1-\epsilon_{veto}) = 7.0\; 10^{-3} {\rm Hz} \;\; (4.3 \; 10^{-3} {\rm Hz} )
\ee

\subparagraph{Muons decaying between the iron shielding and veto}
The last possibility is that the muon decays in the region between the shielding and the veto, mainly  in the very last layer of iron. 
The rate  R$^{\mu\;between\;dec}\sim$ 0.7 Hz\footnote{This rate has been estimated considering  the fraction of $\mu$ decaying in the last 10cm of iron all around the detector  corresponding to few absorption lengths of 20-50 MeV gammas.}.
For a typical decay, the few  $e^+$ that escape the iron have a good chance of being  detected by the veto and fully absorbed by the lead shielding between the veto and CORMORINO; therefore their contribution to the background is negligible.
The rare  $\mu \to e^- \bar\nu_\mu \nu_e \gamma$ decay  (BR$\sim 1.5 \%$) produces a 20-50 MeV gamma that can enter in CORMORINO bypassing the veto. The absorption length for 10 MeV - 100 MeV gammas in lead is a.l.$\sim$0.9cm.  The leads shielding between the detector and the veto is   5 cm thick, corresponding  to 5.5 a.l. for an overall attenuation of $\sim$~4~10$^{-3}$.  The fraction of  10-100 MeV photons interacting with CORMORINO and releasing an energy over threshold is estimated to be: $ \alpha_{Single\; hit}^{\mu\;between\;dec}\sim$0.5 for a 1 MeV threshold and 0.2 at 10 MeV. Considering all these factors, the background rate is 
\be
R_{Bg}^{\mu;\ rare\;dec} = R^{\mu\;between \;dec}\; BR\; Att \; \alpha_{Single\; hit}^{\mu\;between\;dec} = 2\; 10^{-5} {\rm Hz} \; \;(0.8\; 10^{-5}  {\rm Hz} )~,
\ee
and is therefore negligible.
=
\begin{table}[htdp]
\caption{Expected rates for signal,  beam-related and beam-related background in CORMORINO detector for two detection thresholds: 1 MeV and 10 MeV.}
\begin{center}
\begin{tabular}{|c|c|c|}
\hline
 & Rate$_{\;Thr=1 {\rm MeV}}$ (Hz/$\mu{\rm A}$))&Rate$_{\;Thr=10 {\rm MeV}}$ (Hz/$\mu {\rm A}$))\\ 
\hline\hline
 $\chi$ detection  - S.I  & 1.0 10$^{-5}$& 1.2 10$^{-6}$\\
 \hline
$\chi$ detection - S.II  & 2.0 10$^{-7}$ & 0.7 10$^{-7}$\\
 \hline\
B-rel $\nu$ & 2.0 10$^{-9}$ & 2.0 10$^{-10}$\\
 \hline
B-rel neutron  & 0  & 0\\
\hline\hline\hline
 & Rate$_{\;Thr=1 {\rm MeV}}$ (Hz)&Rate$_{\;Thr=10 {\rm MeV}}$ (Hz)\\
\hline\hline
B-unrel $\nu$  &2.0 10$^{-6}$ & 2.0 10$^{-7}$\\
\hline
B-unrel neutron  &2.7 10$^{-3}$ &0.6 10$^{-3}$\\
\hline
Crossing muons  &3.3 10$^{-3}$ &3.5 10$^{-3}$\\
\hline
Captured  $\mu^+$  &1.4 10$^{-3}$  &2.4 10$^{-3}$\\
\hline
Decaying $\mu^-$ (CORM)  &2.9 10$^{-3}$  &4.8 10$^{-3}$\\
\hline
Stopped $\mu$ in lead  &7.0 10$^{-3}$  &4.3 10$^{-3}$\\
\hline
$\mu^-$ rare decay &2.0 10$^{-5}$  &8.0 10$^{-6}$\\
\hline\hline
Total Beam-unrelated bg &1.7 10$^{-2}$  &1.5 10$^{-2}$\\
\hline\hline
\end{tabular}
\end{center}
\label{tab:rates}
\end{table}%

\subsubsection{Summary}
Rates for  signal,   beam-related and beam-unrelated background sources are reported in Tab.~\ref{tab:rates} for the two detection thresholds of 1 MeV and 10 MeV. 
Based on these estimates we can conclude that beam-related backgrounds are negligible compared to cosmogenic sources, which represent the dominant source.
This leads  us to conclude that a test-run for this experimental set up would not give us relevant information because of the limited luminosity.
However, these results call for a thorough investigation of beam-unrelated backgrounds using CORMORINO.
The rates in the table will be  used in the next section to provide an estimate of the   sensitivity for a full scale experiment, with 1 m$^3$ detector for a 6 months run at 100$\mu$A.

\section{Towards a full scale experiment: BDX}\label{sec:fullexp} 
In this Section we present count estimates obtained by scaling-up the rates reported  above
to the full scale experimental setup.
Detailed MC simulations, fully implementing the geometry of an optimized detector and considering the alternative options for beam current, beam-dump geometry,  location, etc. are in progress. Final results will be presented in a full proposal to the next JLab PAC. 

\subsection{The  full experiment}
Here we  consider an experimental set-up that takes advantage of the maximum beam  current available at JLab ($\sim100 \mu$A in Hall-A or Hall-C) assuming that the beam-dump shielding effect will be
comparable to what is estimated  for Hall-D and that the beam-related backgrounds scale correspondingly. We assume to use 1 m$^3$ detector sharing  the same technology of CORMORINO (segmented plastic scintillator bars and PMTs read-out).

Rates for $\chi$-p elastic scattering signal and backgrounds used for this estimate are taken from Tab.~\ref{tab:rates}.
As reported in Sec.~\ref{sec:elastic}, a $\chi$ can also scatter off an atomic electron of the detector. Even if  sub-dominant respect the genuine $\chi$-p elastic scattering, this process can be easily accessed experimentally since it results in a  $\sim$GeV  scattered electron as shown in Fig.~\ref{fig:energyE}.
In the full scale detector,  it will be possible to space out foils of $\sim$mm-thick lead between few cm-thick plastic scintillator bars to build a sampling calorimeter with good energy
resolution and high efficiency. All the considerations and estimates of $\chi$-p elastic detection and the background rates remain valid also with this design. 
Assuming a 15 radiation length calorimeter, it is possible to  obtain a sampling fraction of 30$\%$~\cite{clas-lac} and increase the threshold to 150 MeV (600 MeV)  for a 500 MeV (2 GeV) scattered electron.
 With such high  thresholds we do not expect to have any counts neither from cosmogenic  nor from beam-related background. High energy electron and positron pairs can also be produced by inelastic DM production (see Sec.~\ref{sec:elastic} for details) which yield a similar signal. 

We derived  the number of EOT to be $N_{EOT}\sim10^{22}$, corresponding to 1 year of  parasitic running with 50$\%$ efficiency and a current of 100 $\mu$A.
We scaled the signal and background yield by  assuming a 1 m$^3$ detector with 30 times larger length and the  same transverse size of CORMORINO. 
The full detector would consist of 270 of the CORMORINO modules (instead of 9)
with the surface perpendicular to the beam kept constant to take advantage of the $\chi$ forward kinematics. 
With this geometry,  the isotropic cosmogenic background also scales by a factor of 30. The same is true for  the PMTs and electronic channels, driving 
the costs to a still reasonable value ($\sim\$$M). We also assumed  a beam-unrelated background reduction factor of 5, considering  a conservative 
 time coincidence window between hits in the detector and  RF signal from the beam. The expected counts are reported in Tab.~\ref{tab:results}. The counts
 in the detector are dominated by beam-unrelated background that, as discussed in Sec.~\ref{sec:r-and-d} can be further reduced with an optimized detector and shield.\\
\begin{table}[htdp]
\caption{Expected counts for 6 months of run time at 100 $\mu$A (corresponding to 10$^{22}$ EOT) for signal,  beam-related and beam-unrelated background 
in a 30$\times$ longer  detector for  1 MeV and 10 MeV detection threshold assuming a background  reduction factor 5$\times$ for time coincidence.}
\begin{center}
\begin{tabular}{|c|c|c|}
\hline
 & Counts$_{\;Thr=1 {\rm MeV} }$& Counts$_{\;Thr=10  {\rm MeV}}$\\
\hline\hline
 $\chi$ detection  - S.I &0.5 10$^6$ $\pm$ 700 & 5.7 10$^4\pm$ 240  \\
 \hline
$\chi$ detection - S.II  & 1.0 10$^4$ $ \pm $ 100 & 3.3 10$^3$ $\pm$ 60\\
 \hline\hline\hline
Beam-rel bg & 100 $\pm$ 10  & 10 $\pm$ 3\\
 \hline\hline
Beam-unrel bg  &1.6 10$^6 \pm$ 1300 &1.4 10$^6 \pm$ 1200 \\
\hline\hline
\end{tabular}
\end{center}
\label{tab:results}
\end{table}%

\subsubsection{BDX expected reach} 
\begin{figure}[t!] 
\center
\includegraphics[width=15cm]{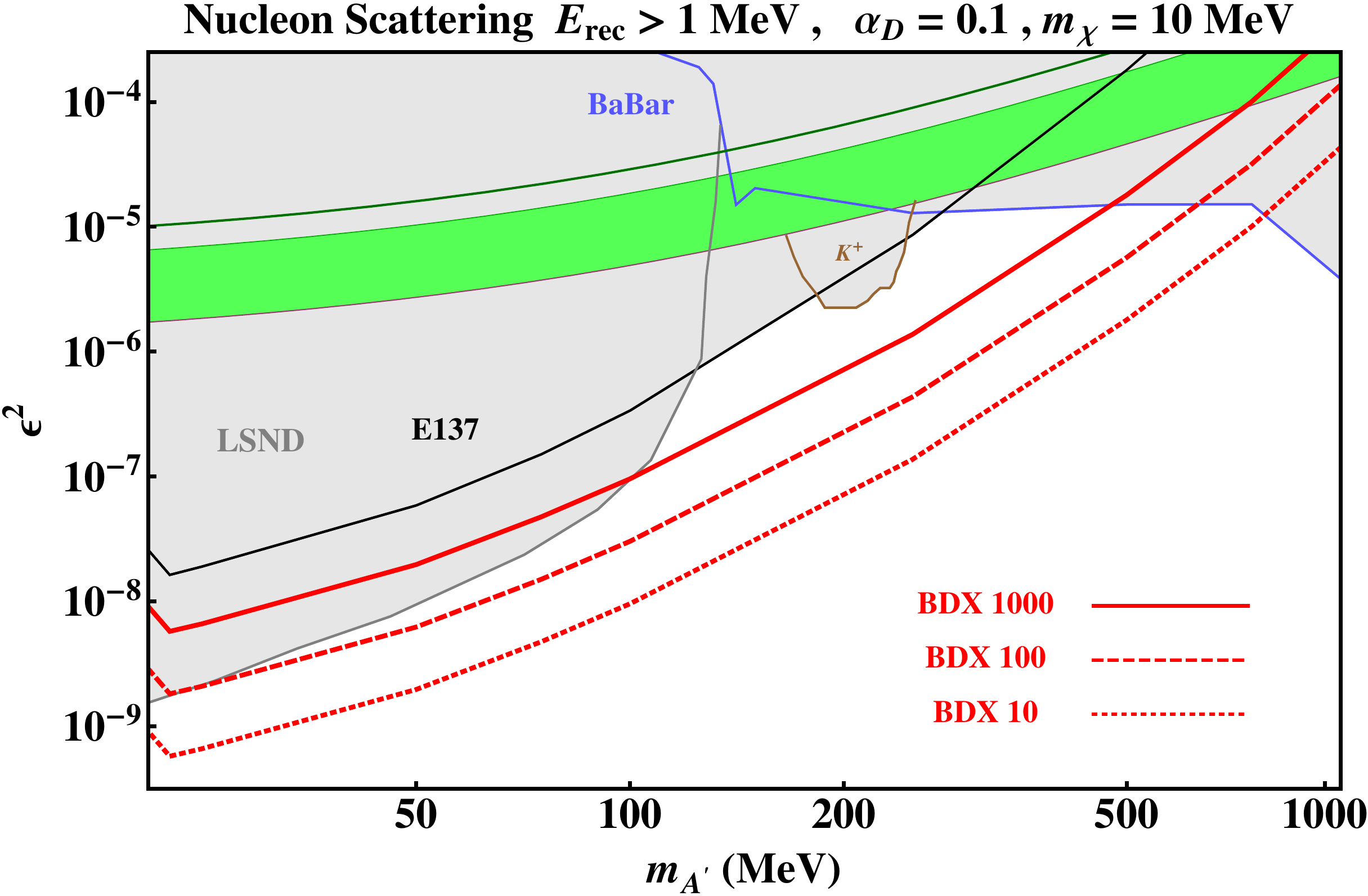} 
 \caption{Red curves show 10, 100, and 1000 event BDX yield projections for a kinetically-mixed dark-photon ($A^\prime$) coupled to a nearly-invisible fermion $\chi$ in the quasi-elastic nucleon recoil channel with 10$^{22}$ EOT. 
 The $\apr$ is radiatively produced in electron-nucleus collisions in the beam-dump and decays promptly to yield $\bar \chi \chi$ pairs, which scatter off detector nucleons and deposit at least 1 MeV of visible energy into the nuclear recoil signal.  
 }
\label{fig:reachN}
 \end{figure}

\begin{figure}[t!] 
\center
\includegraphics[width=0.49\textwidth]{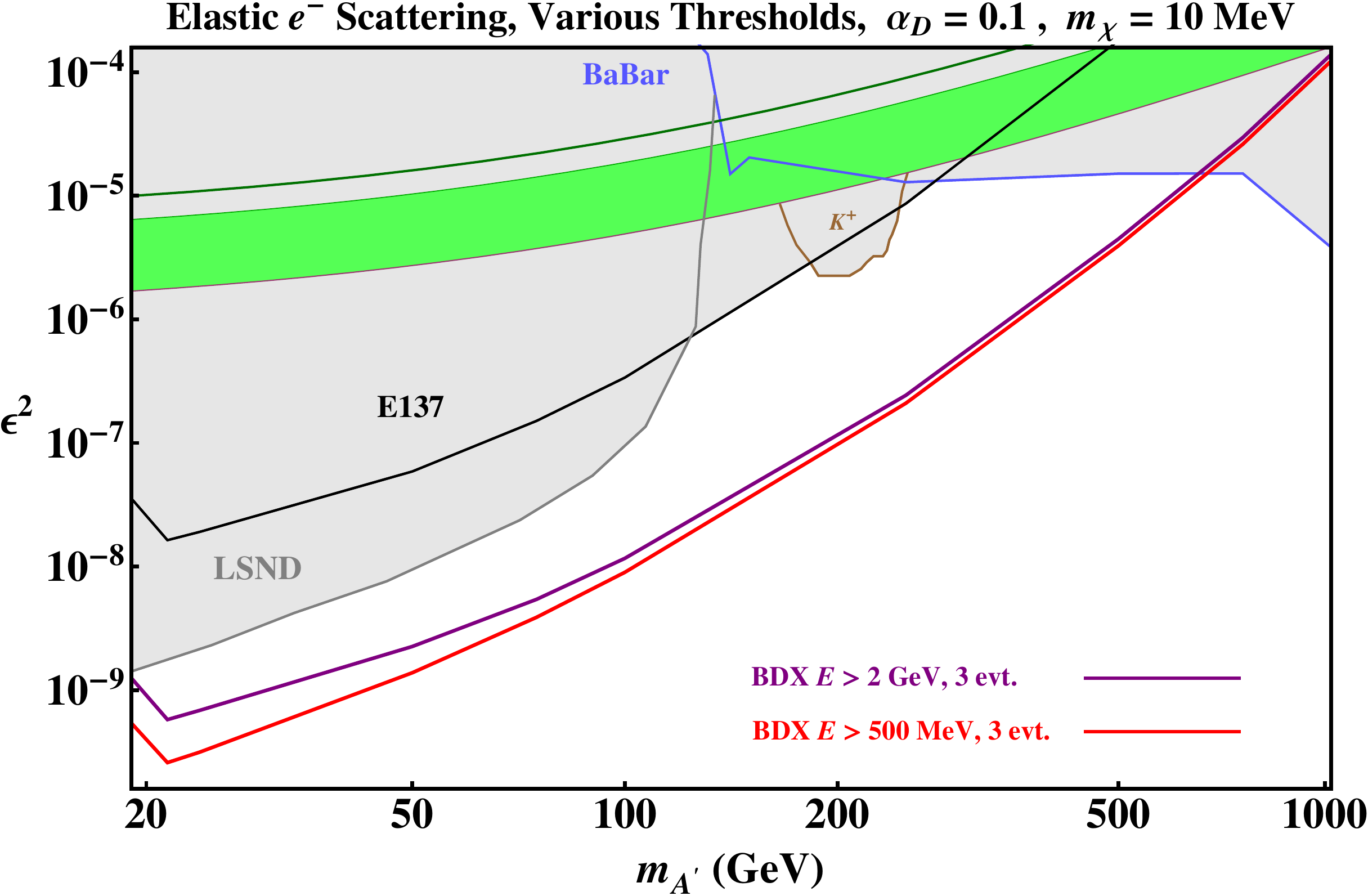} 
\includegraphics[width=0.478\textwidth]{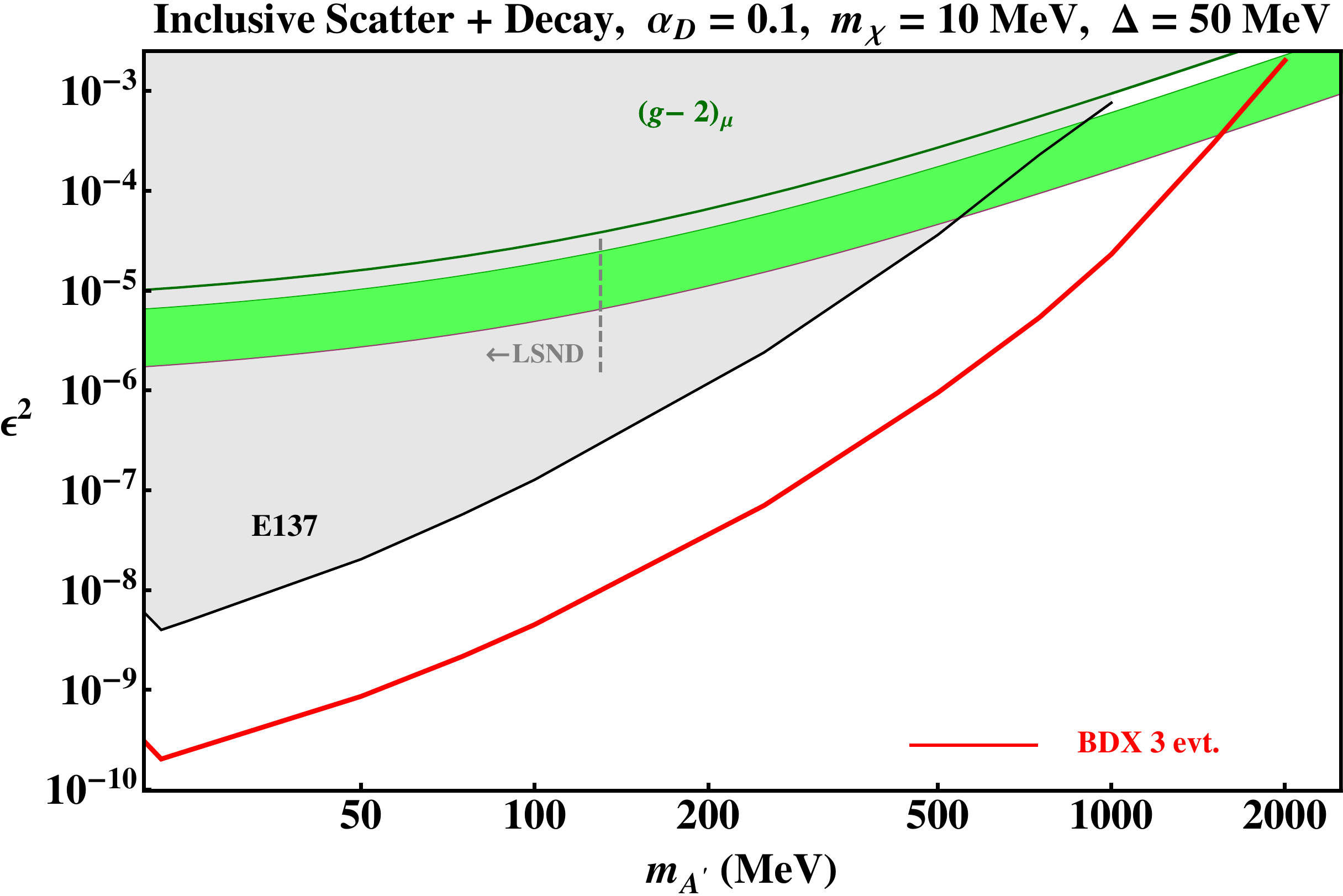}
 \caption{Left: BDX yield projections for a a kinetically-mixed dark-photon ($A^\prime$) coupled to a nearly-invisible Dirac fermion $\chi$ in the electron recoil channel. The experimental setup is identical to that of the previous figure. 
The red and purple curves show 3 event yields with 500 MeV and 2 GeV cuts on electron recoils respectively.
Right:  BDX yield projections for a kinetically mixed dark-photon that decays to two distinct fermions (one lighter, one heavier) whose masses are split by $\Delta = 50$ MeV. 
This interaction arises generically whenever both Dirac and Majorana masses are present in the dark sector. The lighter state can scatter off detector electrons, nucleons, and nuclei to induce an inelastic transition into the excited (heavier) state, which can promptly decay down to the lighter state while simultaneously emitting a {\it visible} electron-positron pair with $\sim$ few GeV of energy.
 The observable in this scenario is the combined electromagnetic energy from this de-excitation including both the target-particle's recoil and energy of the electron positron pair. 
}
\label{fig:reachE} 
\end{figure}

Table~\ref{tab:results} reports fluctuation of backgrounds in the range of 1000 counts for 1 year measurement (50$\%$ efficiency). Based on 
these estimates, a $\chi$-p elastic scattering signal compatible with the least favorable scenario (S.II) 
 would be identified with a significance of 3 $\sigma$ and 10 $\sigma$  over the background for the 1 MeV and 10 MeV threshold, respectively. 
In the event of a null observation, the accumulated data would provide very stringent limits in the DM parameters space. Figures~\ref{fig:reachN} and \ref{fig:reachE} show the BDX sensitivity to the kinetic mixing models of \S\ref{sec:elastic} in the nucleon-recoil and electron channels, respectively. The exclusion regions shown in the plots were derived by  constraints form cosmological observations, meson factories results and  previous electron/hadron  beam-dump experiments. As explained in Sec.~\ref{sec:theory},  limits imposed by E137 experiment  on $\chi$-p elastic scattering were derived within a certain model from $\chi$-e elastic exclusion plot, being the experiment only able to detect multi-GeV electrons. BDX has the unique capability of  being simultaneously sensitive to different DM interaction mechanisms.
The region potentially covered by JLab would, therefore, significantly extend the parameter space covered by previous experiments. 

\subsubsection{Systematic checks} \label{sec:systematics}
In the event of a positive signal, there is a list of possible checks that can be performed to confirm that any observed excess of counts is attributable to a genuine signal.
\begin{itemize}
\item {\bf Beam on/off:} the most straightforward check will be to accumulate data with beam off to determine the cosmogenic background on site; 
assuming the aforementioned 50$\%$ efficiency of running, this corresponds to 6 months of beam-off time.
\item {\bf Beam-related neutrons:}  for half of the run time a lead+water shield  will be placed in the direction of the beam to guarantee  that the count excess is not related to neutrons (scattering and/or captured).
\item {\bf Beam-related neutrinos:} according to  MC simulations, the rate is so low that  even if we scale the results by a factor 100 the neutrino background is still negligible.  Furthermore, we can check the systematic effect by an off-axis measurement since neutrinos are mainly produced isotropically from decays at rest.
\item{ \bf Other beam-related noise:}  due to the expected forward-peaked kinematics of the $\chi$ an off-axis ($\sim 1$m) measurement will verify that the detected signal is due electron beam interactions.
\item{\bf Cosmic background:} a precise measurement of the cosmic background in the detector will be possible and can accumulate data 
either before or after the 1 year experiment time. 
Cosmic background can be measured elsewhere as long as the shielding is identical. In this way a more precise subtraction of the cosmic background will be possible. 
\end{itemize} 

\subsection{Planned R$\&$D}\label{sec:r-and-d}
Results show that, running at one of JLab's high intensity Hall (A or C) beam-dumps with an upgraded version of an existing CORMORINO-like detector, it is possible to explore  a wide range of  model parameters in a time scale of 1 year.  We believe that there is still room for improvement in the detector concept,  active veto, and  shielding design. Moreover, the expected low counting rate coupled with the signal digitization of fADCs will allow us to optimize the off-line reconstruction algorithms, further  enhancing the rejection capability.

As short-term R$\&$D activity, we are planning to validate the simulation results by setting up the veto counter, iron and lead shielding around the CORMORINO detector and measure the cosmogenic background identifying the different sources. 
We are also discussing the option of proposing  a short test run with CORMORINO behind the Hall-D beam-dump before the full detector will be built.
This measurement can help to  further test and validate our detector concept and simulations (e.g. using
the time coincidence for cosmogenic background rejection, as well to  confirm estimates for the beam-related background).

On a mid-term scale, we will investigate other detector options (or upgrades) to reduce costs and increase sensitivity.  Indeed,
a plastic scintillator detector, such as CORMORINO, offers a good initial concept for an optimized detector for a full beam-dump experiment.
It presents many  key-features that we want to preserve in the final detector design:
\begin{itemize}
\item   a high number of  recoiling protons in $\chi-p$ interactions release visible energy in the active target;
\item fast timing ($\sigma_T<70-100$ps) to impose a tight time-coincidence window with a reference signal from the beam (RF) for cosmogenic background rejection;
\item simplicity and reliability of operations -- plastic is an inert material, which remains reasonably stable over time;
\item scalability to larger instrumented volume (at least from CORMORINO  to $\sim$1 m$^3$ size);
\item segmentation for interaction-vertex reconstruction and  background rejection enhancement;
\item easy sensitivity extension to GeV-energy electrons by sampling the electromagnetic shower produced in thin foils of lead placed between the scintillator bars.
\end{itemize}
On the other hand, plastic scintillator presents limitations that may be necessary to overcome in a full scale detector:
\begin{itemize}
\item limited  electron/proton discrimination that, if improved  could be useful to reduce backgrounds;
\item lack of directionality information to correlate hits with a beam-related event;
\item cost of PMTs and number of instrumented channels.
\end{itemize}
We are considering other options such as:  replacing the plastic scintillator with liquid, that, while having similar features, could provide additional advantages;  replacing photo-detectors using SiPM in place of  PMTs; reducing  segmentation to decrease costs. We are also considering  extruded plastic  with fiber/sipm readout to  scale to a multi m$^3$ detector. We are investigating the pros and cons of  each experimental Hall at JLab, evaluating expected reach, logistic,  costs of  civil engineering, and other services.\\ 
Both the final detector design and experimental set up will be discussed in a full proposal to be submitted to the  the next JLab PAC.

\section{Summary and Conclusion}
We propose to mount an experiment behind one of the high-intensity experimental halls at JLab to search for weakly interacting particles produced in
the beam dump. This experiment will have strong, unprecedented sensitivity to dark matter in the MeV -- GeV mass range, probing a parameter space 
two orders of magnitude beyond the reach of past, existing, and proposed experiments. 
 Searches for particles in this mass range are motivated by models that  feature a dark matter particle $\chi$ whose  interactions with the Standard Model (SM) 
 through a new massive dark photon generically appear with strength $\epsilon$ near $10^{-4}-10^{-2}$.
  Such models can also explain the persistent $3 - 4\sigma$ discrepancy between theoretical predictions and experimental 
  observations of the muon's anomalous magnetic moment.

The experiment would detect the interaction of elastically scattered $\chi$s 
off nucleons and electrons 
in a detector situated about 15 m from the beam dump by measuring the
proton and electron recoil energies.  
The sensitivity of the beam dump experiment and its ability to reject backgrounds was studied based on  a case study for an existing detector,
CORMORINO, using detailed modeling of both beam related and cosmogenic backgrounds with a GEANT4 simulation.
The results of this study were scaled to a  1 m$^3$ detector to estimate the sensitivity of a full-scale experiment after collecting $10^{22}$ electrons on target. 
In the absence of a signal and with energy 
thresholds as low as 1 MeV for detecting $\chi-p$ interactions, the experiment would be 
able to set limits on the production of dark matter  with masses in the range between 100 and 500 MeV and coupling constants $\epsilon^2$ between 
10$^{-7}$ and 10$^{-5}$, dominated by statistical uncertainty of cosmogenic backgrounds. 
Elastic $\chi-e$ interactions with thresholds between 0.5 and 2 GeV and essentially no background can be used to set limits on the production of dark 
matter  with masses in the range between 20  and 700 MeV and coupling constants $\epsilon^2$ between  10$^{-9}$ and 10$^{-5}$.
These regions of masses and coupling strengths exceed the expected sensitivity of previous, existing, and proposed experiments by over two orders of magnitude.

\end{document}